\documentclass[11pt]{article}
\usepackage{latexsym}
\usepackage{amssymb}
\usepackage[dvips]{graphicx}
\usepackage{epsfig}
\usepackage{rotating}
\usepackage{amsfonts}
\usepackage{amsmath}
\begin{document}

\begin{titlepage}
\begin{flushright}
CP3-21-45\\
\end{flushright}

\vspace{5pt}

\begin{center}

{\Large\bf Noether Symmetries, Dynamical Constants of Motion,}\\

\vspace{7pt}

{\Large\bf and Spectrum Generating Algebras}\\

\vspace{40pt}

Daddy Balondo Iyela$^a$ and
Jan Govaerts$^{b,c,}$\footnote{Fellow of the Stellenbosch Institute for Advanced Study (STIAS), Stellenbosch,
Republic of South Africa}

\vspace{30pt}

$^{a}${\sl D\'epartement de Physique, Universit\'e de Kinshasa (UNIKIN),\\
B.P. 190 Kinshasa XI, Democratic Republic of Congo}\\
E-mail: {\em balondo36@gmail.com}\\
ORCID: {\tt http://orcid.org/0000-0001-8282-7914}\\
\vspace{15pt}
$^{b}${\sl Centre for Cosmology, Particle Physics and Phenomenology (CP3),\\
Institut de Recherche en Math\'ematique et Physique (IRMP),\\
Universit\'e catholique de Louvain (UCLouvain),\\
2, Chemin du Cyclotron, B-1348 Louvain-la-Neuve, Belgium}\\
E-mail: {\em Jan.Govaerts@uclouvain.be}\\
ORCID: {\tt http://orcid.org/0000-0002-8430-5180}\\
\vspace{15pt}
$^{c}${\sl International Chair in Mathematical Physics and Applications (ICMPA--UNESCO Chair)\\
University of Abomey-Calavi, 072 B.P. 50, Cotonou, Republic of Benin}\\

\vspace{10pt}


\vspace{10pt}

\begin{abstract}
\noindent
When discussing consequences of symmetries of dynamical systems based on Noether's first theorem, most standard textbooks on classical or quantum mechanics
present a conclusion stating that a global continuous Lie symmetry implies the existence of a time independent conserved Noether charge which is the
generator of the action on phase space of that symmetry, and which necessarily must as well commute with the Hamiltonian. However this need not be so,
nor does that statement do justice to the complete scope and reach of Noether's first theorem. Rather a much less restrictive statement applies, namely that
the corresponding Noether charge as an observable over phase space may in fact possess an explicit time dependency, and yet define a constant of the motion
by having a commutator with the Hamiltonian which is nonvanishing, thus indeed defining a dynamical conserved quantity. Furthermore, and this certainly within
the Hamiltonian formulation, the converse statement is valid as well, namely that any dynamical constant of motion is necessarily
the Noether charge of some symmetry leaving the system's action invariant up to some total time derivative contribution. The present contribution
revisits these different points and their consequences, straightaway within the Hamiltonian formulation which is the most appropriate for such issues.
Explicit illustrations are also provided through three general but simple enough classes of systems.
\end{abstract}

\end{center}

\end{titlepage}

\setcounter{footnote}{0}

\section{Introduction}
\label{Intro}

The first Noether theorem, relevant to global continuous Lie symmetries of a dynamical system, is standard textbook material (such discussions seldom include
the second and third Noether theorems as well, relevant to gauge or local continuous Lie symmetries; see for instance Ref.\cite{Gov1}).
Usually, in physics textbooks, the presentation proceeds from the variational principle considered within the Lagrangian formulation for the system's dynamics,
with a regular Lagrange function $L(q^\alpha,\dot{q}^\alpha)$ which is defined over configuration space spanned by the coordinates $q^\alpha(t)$,
with of course $\dot{q}^\alpha(t)=dq^\alpha(t)/dt$. Often this Lagrangian is taken not to have any explicit time dependency in the time evolution coordinate $t$,
and is assumed to be such that the differential $(dt\,L(q^\alpha,\dot{q}^\alpha))$ be exactly  invariant under some class of global continuous transformations
in the variables $q^\alpha$ (or separately in $t$). Consequently such transformations correspond to some form of geometrical symmetry in the space $q^\alpha$
under which the configuration space Euler-Lagrange equations of motion are covariant, thus indeed inducing a mapping between solutions to these equations,
namely a symmetry of these equations leaving invariant the set of their solutions.

As is well known, Noether's first theorem then implies the existence of conserved Noether charges or constants of the motion,
in one-to-one correspondence with each of the independent constant symmetry transformation parameters. These Noether charges are defined
over the associated phase space formulation of the same dynamics spanned by the coordinates $(q^\alpha,p_\alpha)$, $p_\alpha$ being the momenta
canonically conjugate to $q^\alpha$. While under such specific circumstances, these charges are such that they commute with the system's canonical
Hamiltonian $H(q^\alpha,p_\alpha)$, namely possess vanishing Poisson brackets with that Hamiltonian for the classical dynamics, and vanishing
quantum commutators with the quantum Hamiltonian for the quantum dynamics (provided the symmetry is not broken by quantum anomalies incurred
through quantisation). Presumably this is also the reason why in quantum mechanics textbooks more often than not a symmetry is taken to be associated
to a quantum observable which necessarily commutes with the quantum Hamiltonian.

However this need not be so. Indeed there are known instances
of dynamical constants of the motion, which even though conserved through time evolution, do not commute with the Hamiltonian
(see for instance Ref.\cite{Victor}). Their conservation is then made possible by having an explicit time dependency for the corresponding Noether
charges, $Q(q^\alpha,p_\alpha;t)$, such that their Hamiltonian equation of motion implies the following relation between that time dependency and,
say in the classical context, their Poisson bracket with the Hamiltonian,
\begin{equation} 
\frac{\partial Q}{\partial t}=-\left\{Q,H\right\},\qquad
\frac{dQ}{dt}=\frac{\partial Q}{\partial t}\,+\,\left\{Q,H\right\}=0.
\end{equation}

The reach of Noether's first theorem is thus more general still than under the specific restricted conditions recalled above. First a set of
equations of motion may possess symmetries beyond geometrical symmetries in configuration space, the latter thus acting on the configuration space coordinates,
on the one hand, and separately on the conjugate momenta only through an induced action, on the other hand. By extending the analysis based
on the variational principle to the first-order Hamiltonian phase space action for the dynamics, symmetries more general than such geometrical transformations
may be brought to the fore, with their own conserved Noether charges on account of Noether's first theorem which is operative now over phase space,
while such symmetries then mix all phase space configuration space and conjugate momenta coordinates into one another and no longer separately as do geometrical
symmetries. Famous examples of such symmetries of the Hamiltonian formulation which are not manifest geometrical symmetries of the Lagrangian formulation are,
the dynamical SO(4) symmetry of the Coulomb-Kepler problem in 3 dimensional euclidean space, and the dynamical SU($d$) symmetry of the spherical symmetric harmonic oscillator in $d$ dimensional euclidean space. The conserved charges of these examples and symmetries however, do commute with the Hamiltonian,
and do not possess an explicit time dependency in phase space.

Furthermore since equations of motion follow from the variational principle up to terms in the varied action that are total time derivatives
({\sl i.e.}, surface terms in time), symmetry transformations under which the equations of motion are covariant may leave the action invariant but only up
to some total time derivative contribution which is induced by the transformation. Thus in actual fact the (Lagrangian or Hamiltonian) action need not be exactly invariant
under a symmetry transformation, but only up to a total differential in time (as is the case, for instance, for the dynamics of the free non-relativistic massive
particle in $d$ dimensional euclidean space under Galilean boost transformations which are certainly symmetries of that system, as part of the full Galilean
symmetry group of Newton's mechanics in inertial frames)\cite{Gov1,Victor}. Of course Noether's first theorem applies to such circumstances as well,
leading again to constants of the motion as conserved Noether charges.

The scope of Noether's original theorems is thus wider than what is usually presented in most textbooks.
In particular, when the global continuous symmetry transformations
of which the parameters are constants involve (configuration space or phase space) functions which carry an explicit time dependency (think of
Galilean boosts, for instance), then necessarily as phase space functions the associated Noether charges also possess an explicit time dependency, and thus
do not commute with the Hamiltonian --- hence the name of dynamical constants of motion. Nevertheless, even under such circumstances any element
of the enveloping algebra of all conserved charges is again a conserved charge, while the set of conserved charges
is closed under Poisson brackets or commutation relations, whether these are dynamical conserved charges or not.

In the context of the quantised system, these observations have the following consequences. As a subset of this Lie algebra of all constants of the motion,
the Lie sub-algebra of those symmetry conserved charges which do not possess any explicit time dependency has vanishing commutators with
the quantum Hamiltonian. Hence the spectrum of energy eigenstates and their degeneracies is organised into irreducible representations of that Lie sub-algebra,
namely more specifically of its primary generators. This is the situation most often considered in the context of quantum mechanics, which is such
that symmetry transformations associated to time independent conserved charges act by mapping into one another those degenerate energy eigenstates
that belong to a same energy level. Note that these symmetries may extend beyond the Lagrangian geometrical ones, as do for instance
the SO(4) or SU($d$) dynamical symmetries of the two examples mentioned above. On the other hand, for dynamical conserved charges
with their explicit time dependency which therefore do not commute with the quantum Hamiltonian, necessarily the action of the associated symmetries
maps between (linear combinations of) energy eigenstates that belong to different energy levels, thereby generating part, if not all, of the complete
Hilbert space of quantum energy eigenstates for the system. In other words, the enveloping Lie algebra of all (dynamical and non-dynamical) conserved charges
of a system is a spectrum generating Lie algebra for that system.
The Lie sub-algebra of the primary generators of the enveloping Lie algebra of all conserved charges is often referred to as a spectrum generating algebra
or a dynamical algebra of the system (even when no explicit time dependency is involved, which may be sometimes source of some confusion).

Thus given Noether's first theorem, and this now in full generality, whenever the first-order Hamiltonian action of a dynamical system is invariant,
up to some total time derivative contribution, under any global continuous transformation acting on phase space, this transformation defines a symmetry
of that dynamics such that its equations of motion are covariant under that transformation, while that transformation itself, in linearised form, is generated through
the Poisson brackets of the phase space coordinates with a conserved Noether charge which, even when displaying an explicit time dependency, determines
a (then dynamical) constant of the motion.

But then what about the status of the converse statement? Namely, could there exist dynamical constants of the motion which, through their Poisson brackets
generate covariant symmetries of the Hamiltonian equations of motion, and yet would not leave the first-order Hamiltonian action invariant up to some
total time derivative? In other words, could there exist non-Noetherian dynamical constants of the motion beyond those whose existence is a consequence
of Noether's first theorem? In spite of statements to that effect in the literature claiming the possibility of non-Noetherian symmetries,
it will shown hereafter that provided one addresses the issue within the Hamiltonian formulation of the dynamics --- which from a number of points
of view may be considered to be more fundamental than the Lagrangian one --- any dynamical constant of the motion indeed generates through its Poisson brackets
a global continuous symmetry of the dynamics and its equations of motion, and corresponds to the Noether charge associated to that symmetry
which indeed leaves the Hamiltonian action invariant up to a total time derivative. Any dynamical constant of the motion is a Noether charge for some
global continuous symmetry acting on phase space, under which the system's Hamiltonian equations of motion are covariant (for another discussion
of such issues, see Refs.\cite{Leach,Leach2}).

One of the purposes of the present contribution is to (re)visit the different points addressed above, specifically within the Hamiltonian formulation,
and in particular for the converse statement to the first Noether theorem. As far as the present authors are aware, such a general discussion is not readily
available in such a form in the literature, and in particular for the latter point. Furthermore and with the same pedagogical aims in mind, the general analysis
is then illustrated with three explicit examples as well, for which all dynamical constants of the motion and their spectrum generating algebras are identified.

In brief, this paper is organised as follows. In view of the above comments Section 2 revisits Noether's first theorem, and its converse statement,
in a streamlined presentation directly within the Hamiltonian formulation and by assuming the least restrictive and most general possible conditions
under which dynamical constants of motion are indeed always Noether symmetry generators and spectrum generating observables, even when these charges
do not commute with the Hamiltonian. Section~3 provides a first and simple illustration of the general discussion for a system with a purely harmonic dynamics.
This is extended to the free relativistic scalar field in Minkowski spacetime in Section~4. Section~5 details the analysis of the classical and quantum
dynamical constants of motion which generate spatial translations and Galilean boosts in euclidean space in the case of a nonrelativistic particle subjected 
to a constant force field, to highlight through an explicit illustration the consistent role and relevance of dynamical conserved observables as symmetry generators 
and spectrum generating operators. Finally some concluding comments are presented in Section 6.

\section{Symmetries and Conserved Hamiltonian Observables}
\label{Sect2}

\subsection{Noether symmetries and conserved observables}
\label{Sect2.1}

Given the context outlined in the Introduction, let us consider the Hamiltonian formulation of an arbitrary dynamical system, whose phase space
is spanned by local canonical coordinates $(q^\alpha,p_\alpha)$ and its time dynamics governed by a Hamiltonian $H(q^\alpha,p_\alpha)$ (taken here
to be time independent), where $\alpha$ denotes a~---~finite or infinite --- discrete and/or continuous set of possibly multi-indices that label the system's degrees
of freedom. Such a dynamics is characterised by a first-order Hamiltonian action principle given by
\begin{equation}
\label{eq:action1}
S[q^\alpha,p_\alpha]=\int dt\,\left[\lambda\dot{q}^\alpha p_\alpha - (1-\lambda)q^\alpha\dot{p}_\alpha  - H(q^\alpha,p_\alpha)\right],
\end{equation}
where the summation convention is in place, while $\lambda\in\mathbb{R}$ is an arbitrary real constant parameter (which accounts for the freedom
available in redefining the action up to a total time derivative contribution without modifying the canonical character of the phase space coordinates
$(q^\alpha,p_\alpha)$; usual choices for a value of $\lambda$ are $\lambda=1,1/2,0$).
The variational principle implies the canonical Hamiltonian equations of motion,
\begin{equation}
\dot{q}^\alpha=\frac{\partial H}{\partial p_\alpha},\qquad
\dot{p}_\alpha=-\frac{\partial H}{\partial q^\alpha},
\end{equation}
which also follow from the Poisson brackets $\dot{q}^\alpha=\{q^\alpha, H\}$ and $\dot{p}_\alpha=\{p_\alpha,H\}$ given the canonical Poisson brackets
\begin{equation}
\{q^\alpha,p_\beta\}=\delta^\alpha_\beta.
\end{equation}
More generally the time evolution of any phase space observable $F(q^\alpha,p_\alpha;t)$ --- even when complex valued --- having possibly an explicit time
dependency, is determined from the classical Hamiltonian equation of motion,
\begin{equation}
\frac{dF}{dt}=\frac{\partial F}{\partial t} + \{F,H\}.
\end{equation}
Its solutions $F(q^\alpha(t),p_\alpha(t);t)$ are uniquely determined once an initial value has been specified, namely $F_0=F(q^\alpha_0,p_{\alpha,0};t_0)$
where $q^\alpha_0=q^\alpha(t_0)$ and $p_{\alpha,0}=p_\alpha(t_0)$.

According to Noether's first theorem, if the action is left invariant possibly up to a total time derivative contribution under a one-parameter continuous
group of transformations then corresponding to a Lie symmetry of the system and its dynamics (which may be a subgroup of a  many parameter symmetry Lie group),
there exists a combination of the equations of motion that reduces to the total time derivative of a specific quantity~---~namely the corresponding
Noether charge~---~which therefore is itself a conserved observable for any solution to the system's equations of motion, whose constant value is function
of the specific parameters that characterise that solution.

Note well that the present analysis is restricted to one-parameter Lie symmetries, and to dynamics generated by a Hamiltonian function which is not explicitly
time dependent (and thus related through a Legendre transformation to a Lagrange function which itself is time independent).
A more general discussion addressing the possibility of time dependent Hamiltonians (and Lagrangians), and the non-abelian Lie algebra structure itself
of the full ensemble of dynamical Noether constants of motion, is not included in the present work.

Consequently, by having trivially a vanishing Poisson bracket with itself the Hamiltonian identifies a first conserved observable as a Noether charge,
which generates the global continuous symmetry of the dynamics under constant and finite translations in the time evolution parameter.
If $t$ parametrises physical time, this Noether charge measures the total energy of the system. For this reason, let us henceforth restrict to transformations
that leave invariant the time evolution coordinate $t$.

Specifically given the above Hamiltonian formulation let us thus consider the most general possible form of possibly time dependent global continuous transformations
of the phase space variables, $q^\alpha$ and $p_\alpha$, dependent on a single constant parameter $\epsilon$, which would leave the first-order Hamiltonian
action invariant up to a total time derivative contribution. Let us represent such transformations in terms of the following expressions,
\begin{equation}
\tilde{q}^\alpha=\tilde{q}^\alpha(q^\alpha,p_\alpha;t|\epsilon),\qquad
\tilde{p}_\alpha=\tilde{p}_\alpha(q^\alpha,p_\alpha;t|\epsilon),
\end{equation}
such that
\begin{equation}
S[\tilde{q}^\alpha,\tilde{p}_\alpha]=S[q^\alpha,p_\alpha]\,+\,\int dt\,\frac{d}{dt}\left(\Lambda(q^\alpha,p_\alpha;t|\epsilon)\right),
\end{equation}
where $\Lambda(q^\alpha,p_\alpha;t|\epsilon)$ is identified implicitly through the variation of the Hamiltonian action under the transformation
and is defined only up to an arbitrary additive constant contribution.
Since under such conditions the variation of $S[\tilde{q},\tilde{p}_\alpha]$ relative to $(\tilde{q}^\alpha,\tilde{p}_\alpha)$ differs from the variation
of $S[q^\alpha,p_\alpha]$ relative to $(q^\alpha,p_\alpha)$ only by a total time derivative contribution, the Hamiltonian equations of motion of the system
are form invariant under the transformations under consideration. In other words these transformations map solutions into one another, leaving the set of solutions
invariant. They thus define a symmetry of the dynamics.

When considering the linearised form of these transformations, to first order in the global symmetry transformation parameter $\epsilon$,
one establishes the following first Noether identity,
\begin{equation}
\delta S=S[q^\alpha + \delta q^\alpha, p_\alpha + \delta p_\alpha]-S[q^\alpha,p_\alpha]=
\int dt\,\frac{d}{dt}\left(\epsilon\,\Lambda(q^\alpha,p_\alpha;t)\right),
\end{equation}
with $\tilde{q}^\alpha=q^\alpha + \delta q^\alpha$ and $\tilde{p}_\alpha=p_\alpha + \delta p_\alpha$, while
\begin{equation}
\delta q^\alpha = \epsilon\,\phi^\alpha(q^\alpha,p_\alpha;t),\qquad
\delta p_\alpha=\epsilon\,\chi_\alpha(q^\alpha,p_\alpha;t),\qquad
\Lambda(q^\alpha,p_\alpha;t|\epsilon)=\epsilon\,\Lambda(q^\alpha,p_\alpha;t).
\end{equation}
Here $\phi^\alpha(q^\alpha,p_\alpha;t)$, $\chi_\alpha(q^\alpha,p_\alpha;t)$ and $\Lambda(q^\alpha,p_\alpha;t)$ characterise the possibly time dependent action
of the linearised symmetry transformation on the phase space variables $(q^\alpha,p_\alpha)$. Note well that even though the functions
$\phi^\alpha(q^\alpha,p_\alpha;t)$, $\chi_\alpha(q^\alpha,p_\alpha;t)$ and $\Lambda(q^\alpha,p_\alpha;t)$ may possess an explicit time dependency which follows from the nature of the symmetry,
such a transformation still corresponds to (the linearised form of) a global symmetry (as opposed to a local or gauge symmetry) since the symmetry parameter
$\epsilon$ is (a (space)time independent) constant. Furthermore the implicitly identified surface term in time, $\Lambda(q^\alpha,p_\alpha;t)$, remains itself
determined only up to an arbitrary additive constant contribution.

A direct analysis of the above Noether identity for $\delta S$, thus valid for any symmetry transformation leaving the Hamiltonian equations of motion form invariant,
namely the following explicit expression valid to first order in $\epsilon$,
\begin{eqnarray}
\label{eq:Noether1}
\delta S &=& \int dt\,\left[\lambda\left(\frac{d}{dt}\delta q^\alpha\right)p_\alpha + \lambda\left(\frac{dq^\alpha}{dt}\right)\,\delta p_\alpha
-(1-\lambda)\delta q^\alpha\left(\frac{dp_\alpha}{dt}\right)-(1-\lambda)q^\alpha\left(\frac{d\delta p_\alpha}{dt}\right)\right. - \nonumber \\
&& \qquad\qquad\qquad\qquad\qquad  \left. - \delta q^\alpha\frac{\partial H}{\partial q^\alpha}
 -\delta p_\alpha\frac{\partial H}{\partial p_\alpha}\right]=\int dt\,\frac{d}{dt}\left[\epsilon\,\Lambda\right],
\end{eqnarray}
which also reduces to the following identity for the variations $\delta q^\alpha$ and $\delta p_\alpha$ that are determined by the symmetry transformation,
\begin{equation}
0 = \int dt\,\left\{\frac{d}{dt}\left[\lambda\,\delta q^\alpha p_\alpha - (1-\lambda)\,q^\alpha \delta p_\alpha - \epsilon\,\Lambda\right]\,-\,
\delta q^\alpha\,\left(\frac{dp_\alpha}{dt}+\frac{\partial H}{\partial q^\alpha}\right)\,+\,
\delta p_\alpha\,\left(\frac{d q^\alpha}{dt}-\frac{\partial H}{\partial p_\alpha}\right)\right\},
\label{eq:Noether1.2}
\end{equation}
then readily leads to the statement of the first Noether theorem within the Hamiltonian formulation, namely,
\begin{equation}
\frac{dQ}{dt}=-\left(\frac{d q^\alpha}{dt}-\frac{\partial H}{\partial p_\alpha}\right)\chi_\alpha\,+\,
\left(\frac{dp_\alpha}{dt}+\frac{\partial H}{\partial q^\alpha}\right)\phi^\alpha,
\end{equation}
with the Noether charge given by the following phase space observable,
\begin{equation}
\label{eq:Noether2}
Q(q^\alpha,p_\alpha;t)=\lambda\,\phi^\alpha(q^\alpha,p_\alpha;t)\,p_\alpha\,-\,(1-\lambda)\,q^\alpha\,\chi_\alpha(q^\alpha,p_\alpha;t)\,-\,\Lambda(q^\alpha,p_\alpha;t).
\end{equation}
Note that the Noether charge is defined up to the same additive constant that remains arbitrary in the identification of the time surface contribution in $\Lambda$,
a fact which of course is fully consistent with the status of a conserved observable for the Noether charge.

Even though the time surface term $\Lambda$ is defined implicitly through the variation of the Hamiltonian action under the symmetry transformation
up to an additive constant contribution, it is possible to establish a series of conditions in the form of differential equations that the function
$\Lambda(q^\alpha,p_\alpha;t)$ has to satisfy in each of its variables. By making explicit in (\ref{eq:Noether1}) all contributions linear in
$\dot{q}^\alpha$, in $\dot{p}_\alpha$, and independent of these, one identifies the following necessary consistency conditions, respectively,
\begin{eqnarray}
\label{eq:conditions1}
\frac{\partial\Lambda}{\partial q^\alpha} &=& \lambda\frac{\partial\phi^\beta}{\partial q^\alpha} p_\beta + \lambda\chi_\alpha
 - (1-\lambda) q^\beta\frac{\partial\chi_\beta}{\partial q^\alpha}, \nonumber \\
\frac{\partial\Lambda}{\partial p_\alpha} &=& \lambda\frac{\partial\phi^\beta}{\partial p_\alpha} p_\beta - (1-\lambda)\phi^\alpha
- (1-\lambda) q^\beta\frac{\partial\chi_\beta}{\partial p_\alpha} ,  \\
\frac{\partial\Lambda}{\partial t} &=& \lambda\frac{\partial\phi^\alpha}{\partial t}p_\alpha - (1-\lambda) q^\alpha\frac{\partial\chi_\alpha}{\partial t}
-\phi^\alpha\frac{\partial H}{\partial q^\alpha} - \chi_\alpha\frac{\partial H}{\partial p_\alpha}.    \nonumber
\end{eqnarray}
Based on these properties one may then establish through explicit calculation that,
\begin{equation}
\frac{\partial Q}{\partial t}=\phi^\alpha\frac{\partial H}{\partial q^\alpha} + \chi_\alpha\frac{\partial H}{\partial p_\alpha},\qquad
\left\{Q,H\right\}=-\phi^\alpha\frac{\partial H}{\partial q^\alpha} - \chi_\alpha\frac{\partial H}{\partial p_\alpha},
\end{equation}
so that the Hamiltonian equation of motion for the Noether charge reads,
\begin{equation}
\frac{dQ}{dt}=\frac{\partial Q}{\partial t} + \left\{Q,H\right\}=0,
\end{equation}
which is indeed the Hamiltonian equation of motion for a conserved observable which carries an explicit time dependency in addition to its phase space dependency,
with in particular then a nonvanishing Poisson bracket with the system's Hamiltonian.
Furthermore from the identities (\ref{eq:conditions1}) it readily follows that
\begin{equation}
\left\{q^\alpha,\epsilon\, Q\right\}=\epsilon\,\phi^\alpha=\delta q^\alpha,\qquad
\left\{p_\alpha, \epsilon\, Q\right\}=\epsilon\,\chi_\alpha=\delta p_\alpha,
\end{equation}
establishing that the Noether charge $Q$ is indeed the observable that generates the linearised transformations of the corresponding symmetry group's
action on the system's phase space.

\subsection{Conserved observables and Noether symmetries}
\label{Sect2.2}

Let us also explicitly establish the converse result to Noether's first theorem. Namely that any conserved observable, even when carrying an explicit time
dependency over phase space, is the generator of a one-parameter symmetry group of the dynamics which also leaves the first-order Hamiltonian action
invariant up to a total time derivative contribution. Let $F(q^\alpha,p_\alpha;t)$ be such a conserved observable,
thus with the property that
\begin{equation}
\frac{dF}{dt}=\frac{\partial F}{\partial t} + \left\{F,H\right\}=0.
\end{equation}
It readily follows from this conservation property in the form of $\partial F/\partial t=-\left\{F,H\right\}$, that,
\begin{eqnarray}
\frac{d}{dt}\frac{\partial F}{\partial q^\alpha}  &=& \frac{\partial}{\partial t}\frac{\partial F}{\partial q^\alpha} +
\left\{\frac{\partial F}{\partial q^\alpha},H\right\} = -\frac{\partial}{\partial q^\alpha}\left\{F,H\right\} + \left\{\frac{\partial F}{\partial q^\alpha},H\right\}
= -\left\{F,\frac{\partial H}{\partial q^\alpha}\right\}, \nonumber \\
\frac{d}{dt}\frac{\partial F}{\partial p_\alpha} &=& \frac{\partial}{\partial t}\frac{\partial F}{\partial p_\alpha} + 
\left\{\frac{\partial F}{\partial p_\alpha},H\right\} = -\frac{\partial}{\partial p_\alpha}\left\{F,H\right\} + \left\{\frac{\partial F}{\partial p_\alpha},H\right\}
= - \left\{F,\frac{\partial H}{\partial p_\alpha}\right\}.
\end{eqnarray}
Furthermore consider now the linearised phase space transformations generated by $F$ with constant parameter $\epsilon$,
\begin{equation}
\delta q^\alpha=\left\{q^\alpha, \epsilon\,F\right\}=\epsilon\,\frac{\partial F}{\partial p_\alpha} \equiv \epsilon\,\phi^\alpha(q^\alpha,p_\alpha;t),\qquad
\delta p_\alpha=\left\{p_\alpha, \epsilon\,F\right\}=-\epsilon\,\frac{\partial F}{\partial q^\alpha} \equiv \epsilon\,\chi_\alpha(q^\alpha,p_\alpha;t),
\end{equation}
which define the quantities $\phi^\alpha$ and $\chi_\alpha$ as relevant to the present discussion (to be distinguished for now from those introduced
in the previous Subsection \ref{Sect2.1}). These transformations then generate the following linearised variations of the Hamiltonian (note that for any
observable $G(q^\alpha,p_\alpha;t)$ one has to linear order for these transformations,
$G(q^\alpha+\delta q^\alpha,p_\alpha + \delta p_\alpha;t)=G(q^\alpha,p_\alpha;t)+\epsilon\left\{G,F\right\}$, as may easily be checked),
\begin{eqnarray}
\label{eq:varH}
H(q^\alpha+\delta q^\alpha,p_\alpha+\delta p_\alpha) &=& H(q^\alpha,p_\alpha)-\epsilon\,\left\{F,H\right\}, \nonumber \\
\frac{\partial H}{\partial q^\alpha}(q^\alpha + \delta q^\alpha,p_\alpha + \delta p_\alpha) &=&
\frac{\partial H}{\partial q^\alpha}(q^\alpha,p_\alpha) - \epsilon\,\left\{F,\frac{\partial H}{\partial q^\alpha}\right\}, \\
\frac{\partial H}{\partial p_\alpha}(q^\alpha + \delta q^\alpha,p_\alpha + \delta p_\alpha) &=&
\frac{\partial H}{\partial p_\alpha}(q^\alpha,p_\alpha) - \epsilon\,\left\{F,\frac{\partial H}{\partial p_\alpha}\right\}. \nonumber
\end{eqnarray}

Given these series of identities let us now consider a specific but otherwise arbitrary solution to the Hamiltonian equations of motion, $(q^\alpha(t),p_\alpha(t))$,
and its linearly transformed configuration $(\tilde{q}^\alpha(t),\tilde{p}_\alpha(t))$ with
\begin{eqnarray}
\label{eq:tilde1}
\tilde{q}^\alpha(t) &=& q^\alpha(t) + \epsilon\left\{q^\alpha,F\right\}=q^\alpha(t)+\epsilon\frac{\partial F}{\partial p_\alpha}(q^\alpha(t),p_\alpha(t);t)
=q^\alpha(t) + \epsilon\,\phi^\alpha(q^\alpha(t),p_\alpha(t);t), \nonumber \\
\tilde{p}_\alpha(t) &=& p_\alpha(t) + \epsilon\left\{p_\alpha,F\right\}=p_\alpha(t)-\epsilon\frac{\partial F}{\partial q^\alpha}(q^\alpha(t),p_\alpha(t);t)
=p_\alpha(t) + \epsilon\,\chi_\alpha(q^\alpha(t),p_\alpha(t);t).\qquad
\end{eqnarray}
To linear order in $\epsilon$ it then follows simply that,
\begin{eqnarray}
\label{eq:tilde2}
\frac{d}{dt}\tilde{q}^\alpha &=& \frac{\partial H}{\partial p_\alpha}-\epsilon\,\left\{F,\frac{\partial H}{\partial p_\alpha}\right\}
=\frac{\partial H}{\partial p_\alpha}(\tilde{q}^\alpha,\tilde{p}_\alpha), \nonumber \\
\frac{d}{dt}\tilde{p}_\alpha &=& -\frac{\partial H}{\partial q^\alpha}+\epsilon\,\left\{F,\frac{\partial H}{\partial q^\alpha}\right\}
=-\frac{\partial H}{\partial q^\alpha}(\tilde{q}^\alpha,\tilde{p}_\alpha).
\end{eqnarray}
In other words, any solution $(q^\alpha(t),p_\alpha(t))$ to the Hamiltonian equations of motion is transformed into another solution to the same equations
of motion under the action on phase space of the transformation generated by the conserved observable $\epsilon F(q^\alpha,p_\alpha;t)$,
however general the latter may be. Any conserved observable $F$ is thus the generator of a one-parameter Lie symmetry group of the system's dynamics.

The same conclusion may also be reached based directly on the transformation of the first-order Hamiltonian action (\ref{eq:action1}) linearised
to first order in $\epsilon$. A direct analysis then finds for that linearised variation,
\begin{equation}
\delta S=\epsilon\,\int\,dt\left[\frac{d}{dt}\left(\lambda\frac{\partial F}{\partial p_\alpha}p_\alpha + (1-\lambda)q^\alpha\frac{\partial F}{\partial q^\alpha}-F\right)\,+\,
\left(\frac{\partial F}{\partial t}+\left\{F,H\right\}\right)\right].
\end{equation}
Hence indeed when $F$ is a conserved observable, through Poisson brackets it generates phase space transformations that leave the action invariant
up to a total time derivative contribution of a quantity $\Lambda$ given by,
\begin{eqnarray}
\Lambda(q^\alpha,p_\alpha;t) &=& \lambda \frac{\partial F}{\partial p_\alpha} p_\alpha + (1-\lambda)q^\alpha\frac{\partial F}{\partial q_\alpha} - F \nonumber \\
&=& \lambda \phi^\alpha(q^\alpha,p_\alpha;t) p_\alpha - (1-\lambda) q^\alpha\chi_\alpha(q^\alpha,p_\alpha;t) - F(q^\alpha,p_\alpha;t).
\end{eqnarray}
When compared to the general expression (\ref{eq:Noether2}) for the Noether charge associated to the symmetry generated by $F$, it is clear
that the conserved observable $F$ is itself the Noether charge of the symmetry that it generates.

There is thus a one-to-one correspondence between all one-parameter global continuous Lie symmetries of a Hamiltonian dynamics and the set of all its
conserved observables, including those that possess an explicit time dependency over phase space and then do not have a vanishing Poisson bracket
(or quantum commutator) with the Hamiltonian. According to Noether's first theorem, any such Lie symmetry under which the first-order Hamiltonian action is invariant up to a total time derivative contribution and thus such that its Hamiltonian equations of motion are covariant, implies the existence of a (possibly time
dependent) conserved observable which is the generator of that symmetry over phase space through Poisson brackets.
But the converse statement is thus true as well, namely that any
(possibly time dependent) conserved observable is the generator of a one-parameter Lie symmetry under which the Hamiltonian equations of motion are covariant
and for which the first-order Hamiltonian action is invariant up to a total time derivative contribution. Namely any dynamical constant of the motion is the generator
of a Noether symmetry, within the Hamiltonian formulation of a dynamical system. All Lie symmetries are Noetherian symmetries.

Furthermore the set of all conserved observables is a Lie algebra under the Poisson bracket structure. Indeed it is obvious that any linear combination with constant
coefficients of conserved observables is again a conserved observable. The same is true for the product of two conserved observables. Hence the set of all conserved
observables, which generate all the symmetries of a dynamical system, is a closed linear algebra. In addition, because of the Jacobi
identity obeyed by Poisson brackets, it may readily be shown that the Poisson bracket of any two conserved observables is again a conserved observable,
whether explicitly time dependent or not. In other words, the algebra of all conserved observables is a Lie algebra as well, namely the Lie algebra of all
the continuous Lie symmetries of the Hamiltonian dynamics of a dynamical system. Of course, the Lie algebra of all conserved observables may possess closed
Lie sub-algebras of constants of motion, explicitly time dependent or not, of a geometrical character or not, of direct physical relevance and interest.

\section{Harmonic dynamics}
\label{Sect3}

As a first illustration of the general discussion above, let us consider a system with a phase space parametrised by the canonical variables
$(q^\alpha,p_\alpha)$ of which the time evolution is governed by the Hamiltonian
\begin{equation}
H(q^\alpha,p_\alpha)=\sum_\alpha\left(\frac{1}{2}p^2_\alpha + \frac{1}{2}\omega^2_\alpha {q^\alpha}^2\right),
\end{equation}
where $\omega_\alpha>0$ are positive real valued angular frequencies. For instance under the rescaling
$(q^\alpha,p_\alpha)\rightarrow (\sqrt{m}q^\alpha,p_\alpha/\sqrt{m})$ with $m>0$ being a mass parameter, and if the rescaled coordinates
$q^\alpha$ are finite in number and correspond to cartesian coordinates in some euclidean space $\mathbb{R}^d$, this system is that of an
anisotropic harmonic oscillator in that configuration space and represented relative to its principal normal mode axes.

Of course such a system hides no secrets whatsoever. Its equations of motion reduce to the harmonic dynamics of each of the degrees of freedom $q^\alpha$,
\begin{equation}
\ddot{q}^\alpha + \omega^2_\alpha\, q^\alpha = 0,\qquad p_\alpha=\dot{q}^\alpha,
\end{equation}
with obvious solutions.

As is well known, in such a situation it is most appropriate to use a complex parametrisation of phase space, corresponding
to Fock algebra generators in the quantum case, defined by,
\begin{equation}
a_\alpha(q^\alpha,p_\alpha)=\sqrt{\frac{\omega_\alpha}{2}}\left(q^\alpha+\frac{i}{\omega_\alpha} p_\alpha\right),\qquad
a^\dagger(q^\alpha,p_\alpha)=\sqrt{\frac{\omega_\alpha}{2}}\left(q^\alpha-\frac{i}{\omega_\alpha} p_\alpha\right),
\end{equation}
which is such that, for the classical system,
\begin{equation}
\left\{a_\alpha,a^\dagger_\beta\right\}=-i\,\delta_{\alpha,\beta},\qquad
H=\sum_\alpha\,\omega_\alpha a^\dagger_\alpha a_\alpha.
\end{equation}
Consequently one has the following equations of motion,
\begin{equation}
\frac{d}{dt}a_\alpha=\left\{a_\alpha, H\right\}=-i\omega_\alpha a_\alpha,\qquad
\frac{d}{dt}a^\dagger_\alpha=\left\{a^\dagger_\alpha,H\right\}=i\omega_\alpha\,a^\dagger_\alpha,
\end{equation}
of which the solutions are
\begin{equation}
a_\alpha(t)=a_{\alpha,0}\, e^{-i\omega_\alpha (t-t_0)},\qquad
a^\dagger_{\alpha,0}(t)=a^\dagger_{\alpha,0}\,e^{i\omega_\alpha(t-t_0)},
\end{equation}
with $a_{\alpha,0}$ being complex valued integration constants (and $a^\dagger_{\alpha,0}$ their complex conjugates) 
specifying the values of $a_\alpha$ (hence of $q^\alpha$ and $p_\alpha$ as well) at $t=t_0$.

Quite obviously the phase space observables $a_\alpha(q^\alpha,p_\alpha)$ and $a^\dagger_\alpha(q^\alpha,p_\alpha)$, which as such have
no explicit time dependency, are not conserved since they have nonvanishing Poisson brackets (or quantum commutators) with the Hamiltonian $H$.
However let us now introduce the following explicitly time dependent phase space observables,
\begin{equation}
A_\alpha(t)\equiv A_\alpha(q^\alpha,p_\alpha;t)=e^{i\omega_\alpha (t-t_0)}\,a_\alpha(q^\alpha,p_\alpha),\quad
A^\dagger(t)\equiv A^\dagger_\alpha(q^\alpha,p_\alpha;t)=e^{-i\omega_\alpha(t-t_0)}\,a^\dagger_\alpha(q^\alpha,p_\alpha),
\end{equation}
in terms of which the original basic parametrisation $(q^\alpha,p_\alpha)$ of phase space is expressed as,
\begin{eqnarray}
q^\alpha &=& \frac{1}{\sqrt{2\omega_\alpha}}\left(e^{-i\omega_\alpha(t-t_0)}\,A_\alpha(t)\,+\,e^{i\omega_\alpha(t-t_0)}A^\dagger_\alpha(t)\right), \nonumber \\
p_\alpha &=& -\frac{i\omega_\alpha}{\sqrt{2\omega_\alpha}}\left(e^{-i\omega_\alpha(t-t_0)} A_\alpha(t)\,-\,e^{i\omega_\alpha(t-t_0)} A^\dagger_\alpha(t)\right).
\label{eq:Sol1}
\end{eqnarray}
By construction the observables $A_\alpha(t)$ and $A^\dagger_\alpha(t)$ are all conserved quantities, since,
\begin{equation}
\frac{d}{dt}A_\alpha=\frac{\partial}{\partial t} A_\alpha+\left\{A_\alpha,H\right\}=0,\qquad
\frac{d}{dt}A^\dagger_\alpha=\frac{\partial}{\partial t}A^\dagger_\alpha + \left\{A^\dagger_\alpha,H\right\}=0,
\end{equation}
while their constant values for solutions to the Hamiltonian equations of motion are given by the integration constants for the time evolution
of $a_\alpha$ and $a^\dagger_\alpha$,
\begin{eqnarray}
A_\alpha(q^\alpha(t),p_\alpha(t);t) &=& A_\alpha(q^\alpha(t_0),p_\alpha(t_0);t_0)=a_{\alpha,0}, \nonumber \\
A^\dagger_\alpha(q^\alpha(t),p_\alpha(t);t) &=& A^\dagger_\alpha(q^\alpha(t_0),p_\alpha(t_0);t_0)=a^\dagger_{\alpha,0}.
\end{eqnarray}
By substitution in the phase space parametrisation (\ref{eq:Sol1}), one then identifies the solution to the Hamiltonian equations of motion in phase space.
Note that the Poisson brackets of these dynamical constants of the motion still take the canonical values,
\begin{equation}
\left\{A_\alpha(q^\alpha,p_\alpha;t),A^\dagger_\beta(q^\alpha,p_\alpha;t)\right\}=-i\,\delta_{\alpha,\beta}.
\end{equation}

Based on this observation it is clear that the complete algebra of all conserved phase space observables that generate the full dynamical symmetry Lie algebra
of this harmonic dynamics, constructed as composite quantities out of its phase space variables $(q^\alpha,p_\alpha)$, is spanned
by the set of all possible monomials comprised of a product of an arbitrary number of $A^\dagger_\alpha(q^\alpha,p_\alpha;t)$ factors, followed by the product
of an arbitrary number of $A_\alpha(q^\alpha,p_\alpha;t)$ factors, written in normal ordered form relative to these Fock algebra generators in the quantum case,
and this for all possible values of $\alpha$. One example of such a conserved observable being simply, of course, the Hamiltonian itself, in the form,
\begin{equation}
H=\sum_\alpha \omega_\alpha A^\dagger_\alpha(q^\alpha,p_\alpha;t)\,A_\alpha(q^\alpha,p_\alpha;t).
\end{equation}
Given this understanding of the full dynamical symmetry Lie algebra of this system, it also becomes clear that in the quantum case, and then considered within
the Schr\"odinger picture, these monomials which are the generators of this dynamical symmetry, by annihilating and creating arbitrary numbers of arbitrary energy quanta of the system, span as well the complete spectrum generating algebra of all energy eigenstates of this dynamics starting with their action
on the Fock vacuum annihilated by $A_\alpha(q^\alpha,p_\alpha;t)$,
namely $|\Omega\rangle$ such that $A_\alpha(q^\alpha,p_\alpha;t)|\Omega\rangle=0$ and $\langle\Omega|\Omega\rangle=1$.
From that point of view let us remark that the conserved observables $A^\dagger_\alpha(q^\alpha,p_\alpha;t)$ and $A_\alpha(q^\alpha,p_\alpha;t)$
are, on their own already, the primary generators of the full spectrum generating algebra, which is indeed the enveloping algebra constructed out
of these primary conserved observables.

Finally note that by restricting this choice of monomials to those of the following bi-linear form in the creation and annihilation operators for all possible values of $\alpha$ and $\beta$,
namely $A^\dagger_\alpha(t)A_\beta(t)$, one generates a closed subalgebra of conserved observables and Lie symmetry generators
of the full dynamical symmetry Lie algebra, to which the Hamiltonian $H$ belongs. In the quantum case the action of this subalgebra is also closed
in the subspace of Hilbert space spanned by all single quanta quantum states $A^\dagger_\alpha(t)|\Omega\rangle$ for all values of $\alpha$,
even though not all these states are degenerate in energy if the frequencies $\omega_\alpha$ are not all equal.

\section{Free Scalar Fields in $d$+1 Dimensional Minkowski Spacetime}
\label{Sect4}

As a second illustration, let us now consider the relativistic Poincar\'e invariant free real Klein-Gordon scalar field of mass $\mu\ge 0$ in a $D=d$+1
dimensional Minkowski spacetime, $\phi(x^\mu)\in\mathbb{R}$, with its dynamics determined from the usual spacetime local Lagrangian action
(in units such that $c=1=\hbar$),
\begin{equation}
\label{eq:L1}
S[\phi]=\int d^Dx^\mu\left[\frac{1}{2}\left(\partial_t\phi\right)^2-\frac{1}{2}\left(\vec{\nabla}_{\vec{x}}\,\phi\right)^2-\frac{1}{2}\mu^2\phi^2 \right].
\end{equation}
In Hamiltonian form we thus have,
\begin{equation}
S[\phi,\pi]=\int\,d^Dx^\mu\left[\partial_t\phi\,\pi\,-\,\left(\frac{1}{2}\pi^2 + \frac{1}{2}\left(\vec{\nabla}_{\vec{x}}\phi\right)^2+\frac{1}{2}\mu^2\phi^2\right)\right]
=\int d^Dx^\mu\left[\partial_t\phi\,\pi\,-\,{\cal H}\right],
\end{equation}
with $\pi(t,\vec{x})=\partial_t\phi(t,\vec{x})$ being the canonical conjugate momentum to $\phi(t,\vec{x})$, and ${\cal H}$ the canonical Hamiltonian density.
It should be clear, by considering the momentum space representation of the field in terms of its Fourier transformed modes, that this system is
of the general harmonic form discussed in the previous Section. The analysis thus readily proceeds accordingly.

Given the original phase space parametrisation in terms of the canonically conjugate variables $(\phi(\vec{x}),\pi(\vec{x}))$, let us now consider its complex parametrisation
in terms of complex variables ``in momentum space", $(a(\vec{k}),a^\dagger(\vec{k}))$, defined through a Fourier transform such that,
\begin{eqnarray}
\phi(\vec{x}) &=& \int_{(\infty)}\frac{d^d\vec{k}}{(2\pi)^d\, 2\omega(k)}\left(e^{i\vec{k}\cdot\vec{x}}\,a(\vec{k})\,+\,e^{-i\vec{k}\cdot\vec{x}}\,a^\dagger(\vec{k})\right), \nonumber \\
\pi(\vec{x}) &=& \int_{(\infty)}\frac{d^d\vec{k}}{(2\pi)^d \, 2\omega(k)}\left(-i\omega(k)\right)
\left(e^{i\vec{k}\cdot\vec{x}}\,a(\vec{k})\,-\,e^{-i\vec{k}\cdot\vec{x}}\,a^\dagger(\vec{k})\right),
\end{eqnarray}
where $\omega(k)=\sqrt{\vec{k}^2+\mu^2}$ and $k=|\vec{k}|$. Note that the plane wave factors involved in this mode decomposition of the field, namely
$e^{i\vec{k}\cdot\vec{x}}/(2\pi)^{d/2}$, meet the necessary completeness relation,
\begin{equation}
\int_{(\infty)}\frac{d^d\vec{k}}{(2\pi)^d}\,e^{-i\vec{k}\cdot\vec{x}}\,e^{i\vec{k}\cdot\vec{y}}=\delta^{(d)}(\vec{x}-\vec{y}\,),
\end{equation}
while they also determine a complete set of orthonormalised functions in space,
\begin{equation}
\int_{(\infty)}\frac{d^d\vec{x}}{(2\pi)^d}\,e^{-i\vec{k}\cdot\vec{x}}\,e^{i\vec{\ell}\cdot\vec{x}}=\delta^{(d)}(\vec{k}-\vec{\ell}\,).
\end{equation}

As is well known, in the context of the quantised field theory already, one then has the following
commutation relations for this Fock algebra parametrisation,
\begin{equation}
\left[a(\vec{k}),a^\dagger(\vec{\ell})\right]=(2\pi)^d\,2\omega(k)\,\delta^{(d)}(\vec{k}-\vec{\ell}\,)\,\mathbb{I},
\end{equation}
and the following expressions for the Hamiltonian, $H$, or total conserved energy, and the total conserved momentum of the quantised field,
given the choice of normal ordering relative to the Fock algebra generators $(a(\vec{k}),a^\dagger(\vec{k}))$ for composite operators, namely,
\begin{equation}
H=\int_{(\infty)}\frac{d^d\vec{k}}{(2\pi)^d\, 2\omega(k)}\,a^\dagger(\vec{k})\,\omega(k)\,a(\vec{k}),\qquad
\vec{P}=\int_{(\infty)}\frac{d^d\vec{k}}{(2\pi)^d\, 2\omega(k)}\,a^\dagger(\vec{k})\,\vec{k}\,a(\vec{k}),
\end{equation}
with indeed,
\begin{equation}
\left[H,a^\dagger(\vec{k})\right]=\omega(k)\,a^\dagger(\vec{k}),\qquad
\left[\vec{P},a^\dagger(\vec{k})\right]=\vec{k}\,a^\dagger(\vec{k}),\qquad
\left[\vec{P},H\right]=0.
\end{equation}

The Heisenberg equations of motion for the Fock algebra operators are,
\begin{equation}
\frac{d}{dt}a(\vec{k})=-i\left[a(\vec{k}),H\right]=-i\omega(k)\,a(\vec{k}),\qquad
\frac{d}{dt}a^\dagger(\vec{k})=-i\left[a^\dagger(\vec{k}),H\right]=i\omega(k)\,a^\dagger(\vec{k}).
\end{equation}
These nonconserved observables thus have the following time dependency for solutions to the quantum dynamics of the system,
\begin{equation}
a(t,\vec{k})=e^{-i\omega(k)(t-t_0)}\,a(t_0,\vec{k}),\qquad
a^\dagger(t,\vec{k})=e^{i\omega(k)(t-t_0)}\,a^\dagger(t_0,\vec{k}),
\end{equation}
where $a(t_0,\vec{k})$ and $a^\dagger(t_0,\vec{k})$ are initial values for these observables at the reference time $t=t_0$.

Thus once again, let us consider the following basic conserved quantum observables possessing an explicit time dependency in phase space,
and related to the Fock algebra at the reference time $t_0$ and considered in the Schr\"odinger picture,
\begin{equation}
A(\vec{k};t)=e^{i\omega(k)(t-t_0)}\,a(\vec{k}),\qquad
A^\dagger(\vec{k};t)=e^{-i\omega(k)(t-t_0)}\,a^\dagger(\vec{k}),
\end{equation}
whose commutation relations remain of the Fock algebra type,
\begin{equation}
\left[A(\vec{k};t),A^\dagger(\vec{\ell};t)\right]=(2\pi)^d\,2\omega(k)\,\delta^{(d)}(\vec{k}-\vec{\ell}\,)\,\mathbb{I}.
\end{equation}
In the Heisenberg picture, these observables are constants of the motion since
$dA_H(\vec{k};t|t)/dt=\partial A_H(\vec{k};t|t)/\partial t -i \left[A_H(\vec{k};t|t),H\right]=0$
(with the partial time derivative in the first term in the r.h.s.~of this expression acting only on the explicit time dependency of the phase space
observable $A(\vec{k};t)$ in the Schr\"odinger picture), namely,
\begin{eqnarray}
A_H(\vec{k};t|t)=A_H(\vec{k};t_0|t_0)=a(t_0,\vec{k}),\qquad
A^\dagger_H(\vec{k};t|t)=A^\dagger_H(\vec{k};t_0|t_0)=a^\dagger(t_0,\vec{k}).
\end{eqnarray}
Furthermore the original phase space variables are then expressed as follows in terms of the phase space parametrisation $(A_H(\vec{k};t|t),A^\dagger_H(\vec{k};t|t))$
in the Heisenberg picture,
\begin{eqnarray}
\phi_H(t,\vec{x})\!\!&=&\!\!\! \int_{(\infty)}\frac{d^d\vec{k}}{(2\pi)^d\, 2\omega(k)}\left(e^{-i\omega(k)(t-t_0)+i\vec{k}\cdot\vec{x}}\,A_H(\vec{k};t|t)
+e^{i\omega(k)(t-t_0)-i\vec{k}\cdot\vec{x}}\,A^\dagger_H(\vec{k};t|t)\right), \\
\pi_H(t,\vec{x})\!\! &=&\!\!\! \int_{(\infty)}\frac{d^d\vec{k}}{(2\pi)^d\, 2\omega(k)}\left(-i\omega(k)\right)
\left(e^{-i\omega(k)(t-t_0)+i\vec{k}\cdot\vec{x}}\,A_H(\vec{k};t|t) - e^{i\omega(k)(t-t_0)-i\vec{k}\cdot\vec{x}}\,A^\dagger_H(\vec{k};t|t)\right), \nonumber
\end{eqnarray}
thereby leading to the usual expression for the time dependency of the dynamical quantum scalar field in the Heisenberg picture,
in terms of the integration constants $(a(t_0,\vec{k}),a^\dagger(t_0,\vec{k}))$. Note that in the Heisenberg picture the conserved total
energy-momentum of the field is expressed as follows in terms of the conserved Fock algebra generators,
\begin{equation}
H=\int_{(\infty)}\frac{d^d\vec{k}}{(2\pi)^d\, 2\omega(k)}\,A^\dagger_H(\vec{k};t|t)\,\omega(k)\,A_H(\vec{k};t|t),\quad
\vec{P}=\int_{(\infty)}\frac{d^d\vec{k}}{(2\pi)^d\, 2\omega(k)}\,A^\dagger_H(\vec{k};t|t)\,\vec{k}\,A_H(\vec{k};t|t).\qquad
\end{equation}

Obviously the full dynamical symmetry Lie algebra and spectrum generating algebra of the system is spanned by the full enveloping algebra generated
by the primary conserved Fock space operators, considered now within the Schr\"odinger picture, namely $A(\vec{k};t)$ and $A^\dagger(\vec{k};t)$.
In particular the sub-algebra of bi-linears in the creation and annihilation operators, $A^\dagger(\vec{k}_1;t)\,A(\vec{k}_2;t)$, acts on the subspace
of Hilbert space comprised of all 1-particle quantum states with definite momentum eigenvalue by changing their momentum and energy values.
Yet, these transformations define global continuous Lie symmetries of the free field dynamics, even though they do not commute with the quantum Hamiltonian
when $\omega(k_1)\ne \omega(k_2)$. Note that comments made to that effect in the Conclusions of Ref.\cite{BG1} thus prove to have been misconceived.

\section{The Nonrelativistic Particle in a Constant Force Field}
\label{Sect5}

As a third and last illustration, let us consider a nonrelativistic particle of mass $m>0$ with position vector $\vec{x}(t)$ relative to some inertial frame
 in a $d$-dimensional euclidean space, subjected to a constant and thus conservative force $\vec{F}$ (see for instance Ref.\cite{Victor}),
 for which the configuration space equation of motion reads,
\begin{equation}
m\frac{d^2\vec{x}}{dt^2}=\vec{F}=-\vec{\nabla}_{\vec{x}} V(\vec{x}\,),\qquad
V(\vec{x}\,)=-\vec{x}\cdot\vec{F}.
\label{eq:particle}
\end{equation}
Geometrical global continuous symmetries of this dynamics include constant translations in time, constant translations in space, constant rotations
in space that leave invariant the force vector $\vec{F}$, and constant velocity Galilean boosts in space. To each class of these transformations
there correspond conserved Noether charges which are the generators of these transformations once the phase space parametrisation of the system
is considered. Knowing that the Hamiltonian is the Noether charge for constant time translations, hereafter the dynamical conserved charges associated
to constant space translations and constant velocity Galilean boosts are being considered specifically, leaving aside the generators for spatial rotations
in those planes perpendicular to $\vec{F}$. Denoting by $\vec{a}$ the constant spatial translation vector, and by $\vec{v}$ the constant Galilean boost vector
(not to be confused with the particle's velocity, $\dot{\vec{x}}(t)$), the action of these transformations on the degrees of freedom $\vec{x}(t)$
(namely $x^i(t)$ with $i=1,2,\cdots,d$) and the velocity momentum $\vec{p}(t)=m\dot{\vec{x}}(t)$ are represented as follows,
\begin{equation}
\vec{x}\,'(t)=\vec{x}(t)\,+\,\vec{a}\,+\,t\,\vec{v},\qquad
\vec{p}\,'(t)=\vec{p}(t)\,+\,m\,\vec{v}.
\end{equation}

\subsection{Classical dynamics and dynamical constants of the motion}
\label{Sect5.1}

The above equation of motion (\ref{eq:particle}) is of course the Euler-Lagrange equation given the following choice of Lagrangian action for this system,
\begin{equation}
S_{\rm L}[\vec{x}]=\int dt\,\left[\frac{1}{2}m\dot{\vec{x}}^{\,2}\,+\,\vec{x}\cdot\vec{F}\right].
\end{equation}
Its associated first-order Hamiltonian action is given as, by explicitly using as well the free parameter $\lambda\in\mathbb{R}$ as in the general discussion
of Sect.\ref{Sect2},
\begin{equation}
S[\vec{x},\vec{p}\,]=\int dt\left[\,\lambda\,\dot{\vec{x}}\cdot\vec{p}\,-\,(1-\lambda)\,\vec{x}\cdot\dot{\vec{p}}\,-\,H(\vec{x},\vec{p}\,)\right],\qquad
H(\vec{x},\vec{p}\,)=\frac{1}{2m}\vec{p}\,^2\,-\,\vec{x}\cdot\vec{F},
\end{equation}
where the canonical conjugate momentum coincides with the velocity momentum, $\vec{p}=m\dot{\vec{x}}$.

Clearly whether in the Lagrangian or the Hamiltonian formulation, none of these actions are exactly invariant under the considered symmetry transformations,
but they are indeed invariant up to a total time derivative contribution which is a necessary and sufficient condition for a Noether symmetry.
In the case of the Lagrangian action one has,
\begin{equation}
S_{\rm L}[\vec{x}\,']=S_{\rm L}[\vec{x}]+
\int dt\frac{d}{dt}\left[m\vec{v}\cdot\vec{x}+\frac{1}{2}t\,m\vec{v}\,^2 + t\, \vec{a}\cdot\vec{F} + \frac{1}{2}t^2\, \vec{v}\cdot\vec{F}\right],
\end{equation}
while the first-order Hamiltonian action is such that
\begin{equation}
S[\vec{x}\,',\vec{p}\,']=S[\vec{x},\vec{p}\,]+\int dt\frac{d}{dt}\Lambda(\vec{x},\vec{p};t|\vec{a},\vec{v}),
\end{equation}
where
\begin{equation}
\Lambda(\vec{x},\vec{p};t|\vec{a},\vec{v})=\lambda\,m\vec{v}\cdot\vec{x} + \lambda t\,m\vec{v}\,^2
-(1-\lambda)\,t\,\vec{v}\cdot\vec{p} - \frac{1}{2}t\,m\vec{v}\,^2 + \frac{1}{2}t^2\,\vec{v}\cdot\vec{F}
-(1-\lambda)\,\vec{a}\cdot\vec{p} + t\,\vec{a}\cdot\vec{F}.
\end{equation}
Therefore in linearised form to first order in $\vec{a}$ and $\vec{v}$ one obtains,
\begin{equation}
\Lambda(\vec{x},\vec{p};t|\vec{a},\vec{v}) \simeq \vec{a}\cdot\left[-(1-\lambda)\,\vec{p} + t\,\vec{F}\right]\,+\,\vec{v}\cdot
\left[\lambda\,m\vec{x}\,-\,(1-\lambda)\,t\,\vec{p}\,+\,m\,\frac{1}{2}t^2\,\vec{F}\right].
\end{equation}

On the other hand, since in the notations of Sect.\ref{Sect2.1} one has for the transformed phase space coordinates,
\begin{equation}
\delta\vec{x}=\vec{a}+t\,\vec{v},\qquad
\delta\vec{p}=m \vec{v},
\end{equation}
and given the identification (\ref{eq:Noether2}) for Noether charges, one readily identifies the Noether charge for spatial translations to read,
\begin{equation}
\vec{T}(\vec{x},\vec{p};t)=\vec{p}\,-\,t\,\vec{F},
\end{equation}
and that for Galilean boosts,
\begin{equation}
\vec{\gamma}(\vec{x},\vec{p};t)=-m\vec{x}+t\,\vec{p}\,-\,\frac{1}{2}t^2\,\vec{F},
\end{equation}
both of which are thus dynamical constants of motion inclusive of their explicit time dependency as phase space observables.
Explicitly one finds for their Hamiltonian equations of motion,
\begin{equation}
\frac{\partial\vec{T}}{\partial t}=-\vec{F},\quad
\left\{\vec{T},H\right\}=\vec{F},\qquad
\frac{\partial\vec{\gamma}}{\partial t}=\vec{p}-t\vec{F}=\vec{T},\quad
\left\{\vec{\gamma},H\right\}=-\vec{p}+t\vec{F}=-\vec{T},
\end{equation}
so that indeed,
\begin{equation}
\frac{d\vec{T}}{d t}=\frac{\partial\vec{T}}{\partial t}+\left\{\vec{T},H\right\}=\vec{0},\qquad
\frac{d\vec{\gamma}}{d t}=\frac{\partial\vec{\gamma}}{\partial t}+\left\{\vec{\gamma},H\right\}=\vec{0}.
\end{equation}

Furthermore the Poisson bracket algebra of these Noether charges is given as,
\begin{equation}
\left\{ T_i,T_j\right\}=0,\quad
\left\{\gamma_i,\gamma_j\right\}=0,\quad
\left\{T_i,\gamma_j\right\}=m\delta_{ij},\quad
\left\{\vec{T},H\right\}=\vec{F},\quad
\left\{\vec{\gamma},H\right\}=-\vec{T},
\end{equation}
showing that the mass parameter $m$ determines a classical central extension of the algebra of the Galilean group\cite{Lucio}, with a value independent
of the external force $\vec{F}$. In particular note how the dynamical observables $T_i(\vec{x},\vec{p};t)$ and $\gamma_i(\vec{x},\vec{p};t)/m$ are
canonically conjugate variables in phase space for all values of $t$, with the same Poisson brackets as the basic phase space coordinates $x_i$ and $p_i$.
In addition, since one finds,
\begin{equation}
\left\{\vec{x},\vec{a}\cdot\vec{T}\right\}=\vec{a},\quad
\left\{\vec{p},\vec{a}\cdot\vec{T}\right\}=\vec{0},\quad
\left\{\vec{x},\vec{v}\cdot\vec{\gamma}\right\}=t\,\vec{v},\quad
\left\{\vec{p},\vec{v}\cdot\vec{\gamma}\right\}=m\,\vec{v},
\end{equation}
the conserved quantities $\vec{T}$ and $\vec{\gamma}$ are indeed the phase space generators of spatial translations and of Galilean boosts, respectively,
themselves transforming as,
\begin{equation}
\left\{\vec{T},\vec{a}\cdot\vec{T}\right\}=\vec{0},\quad
\left\{\vec{\gamma},\vec{a}\cdot\vec{T}\right\}=-m\vec{a},\quad
\left\{\vec{T},\vec{v}\cdot\vec{\gamma}\right\}=m\vec{v},\quad
\left\{\vec{\gamma},\vec{v}\cdot\vec{\gamma}\right\}=\vec{0}.
\end{equation}

The $(\vec{x},\vec{p}\,)$ phase space parametrisation may also be expressed in terms of these Noether charges, as,
\begin{equation}
\vec{p}=\vec{T}(\vec{x},\vec{p};t) + t\,\vec{F},\qquad
\vec{x}=\frac{1}{m}\left(-\vec{\gamma}(\vec{x},\vec{p};t) + t\,\vec{T}(\vec{x},\vec{p};t)+\frac{1}{2}t^2\,\vec{F}\right).
\end{equation}
In particular, since the solution to the Hamiltonian equations of motion for the Noether charges is trivial and in the form of,
\begin{equation}
\vec{T}(\vec{x}(t),\vec{p}(t);t)=\vec{T}(\vec{x}_0,\vec{p}_0;0)=\vec{p}_0,\qquad
\vec{\gamma}(\vec{x}(t),\vec{p}(t);t)=\vec{\gamma}(\vec{x}_0,\vec{p}_0;0)=-m\vec{x}_0,
\end{equation}
where $\vec{x}_0$ and $\vec{p}_0$ are the initial values for $\vec{x}(t)$ and $\vec{p}(t)$ at $t=0$, respectively, the solution to the Hamiltonian
equations of motion for the latter degrees of freedom is readily given as, directly from their representation above in terms of the Noether charges,
\begin{equation}
\vec{x}(t)=\vec{x}_0 + \frac{t}{m}\vec{p}_0+\frac{t^2}{2m}\vec{F},\qquad
\vec{p}(t)=\vec{p}_0 + t \vec{F},
\end{equation}
of course as it should.

Consequently the full dynamical symmetry Lie algebra of the system is simply the enveloping algebra of the dynamical conserved Noether charges
$\vec{T}$ and $\vec{\gamma}$, which are the primary generators of that algebra also as a spectrum generating algebra. For instance in terms of these
constants of the motion the Hamiltonian of the particle reads,
\begin{equation}
H(\vec{x},\vec{p}\,)=\frac{1}{2m}\vec{T}\,^2(\vec{x},\vec{p};t)+\frac{1}{m}\vec{F}\cdot\vec{\gamma}(\vec{x},\vec{p};t),
\end{equation}
with the following constant value for the solution determined by the initial conditions $\vec{x}_0$ and $\vec{p}_0$,
\begin{equation}
H(\vec{x}(t),\vec{p}(t))=\frac{1}{2m}\vec{p}_0\,^2 - \vec{x}_0\cdot\vec{F}.
\end{equation}
This observation is also consistent with the variation of the Hamiltonian under finite transformations generated by these two Noether charges,
given as,
\begin{equation}
H(\vec{x}+\vec{a}+t\vec{v},\vec{p}+m\vec{v}\,)\,-\,H(\vec{x},\vec{p}\,)=
-\vec{a}\cdot\vec{F} + \vec{v}\cdot\vec{T} + \frac{1}{2}m\vec{v}\,^2=
-\vec{a}\cdot\vec{F} + \vec{v}\cdot\left(\vec{p}-t\vec{F}\right)+\frac{1}{2}m\vec{v}\,^2,
\end{equation}
a result which coincides with the finite second order recursive action of the Noether charges through Poisson brackets, namely,
\begin{eqnarray}
&&H(\vec{x}+\vec{a}+t\vec{v},\vec{p}+m\vec{v}\,) = H(\vec{x},\vec{p}\,)+
\left\{H(\vec{x},\vec{p}\,),\vec{a}\cdot\vec{T}(\vec{x},\vec{p};t)+\vec{v}\cdot\vec{\gamma}(\vec{x},\vec{p};t)\right\} + \nonumber \\
&&\qquad\qquad +\frac{1}{2}\left\{\left\{H(\vec{x},\vec{p}\,),\vec{a}\cdot\vec{T}(\vec{x},\vec{p};t)+\vec{v}\cdot\vec{\gamma}(\vec{x},\vec{p};t)\right\},
\vec{a}\cdot\vec{T}(\vec{x},\vec{p};t)+\vec{v}\cdot\vec{\gamma}(\vec{x},\vec{p};t)\right\}.
\end{eqnarray}
Thus spatial translations and Galilean boosts, which are global continuous Lie symmetries of the dynamics, indeed map between solutions with different energy values,
since their generators do not commute with the Hamiltonian even though these observables constitute the primary generators of all dynamical conserved
quantities of the system.

Incidentally note that the Noether charges of the remaining geometrical symmetry not included in the present discussion, namely SO($d-1$) spatial rotations
in all those planes which are perpendicular to $\vec{F}$, already belong to the enveloping algebra generated by the Noether dynamical charges
$\vec{T}$ and $\vec{\gamma}$. Indeed if the components of $\vec{x}$ and $\vec{p}$ that are perpendicular to $\vec{F}$ are denoted by
$\vec{x}_\bot$ and $\vec{p}_\bot$, respectively, and likewise for those of $\vec{T}$ and $\vec{\gamma}$,
the Noether charges for these spatial rotations are given in the following form,
\begin{equation}
L_{ij}=x_{\bot,i}\, p_{\bot,j} - x_{\bot,j}\, p_{\bot,i}=-\frac{1}{m}\left(\gamma_{\bot,i}\,T_{\bot,j}\,-\,\gamma_{\bot,j}\,T_{\bot,i}\right).
\end{equation} 

\subsection{Quantising the dynamics --- at $t=0$}
\label{Sect5.2}

Given the above canonical Hamiltonian formulation of the system, let us consider its quantisation relative to the reference time $t=0$. The space of quantum states
is a representation of the commutation relations for the basic phase space observables at $t=0$, promoted now to quantum operators
$(\hat{\vec{x}}_0\equiv\hat{\vec{x}}(t=0),\hat{\vec{p}}_0\equiv\hat{\vec{p}}(t=0))$ which span a tensor product
of Heisenberg algebras, namely (only the nonvanishing commutators are given),
\begin{equation}
\left[\hat{x}_{0,i},\hat{p}_{0,j}\right]=i\hbar\,\delta_{ij}\,\mathbb{I},\qquad
\hat{\vec{x}}_0\,^\dagger=\hat{\vec{x}}_0,\qquad
\hat{\vec{p}}_0\,^\dagger=\hat{\vec{p}}_0.
\end{equation}
The quantum Hamiltonian is then expressed as,
\begin{equation}
\hat{H}=\frac{1}{2m}\hat{\vec{p}}_0^{\,2}\,-\,\hat{\vec{x}}_0\cdot\vec{F},
\end{equation}
which is trivially such that
\begin{equation}
\frac{\partial \hat{H}}{\partial t}=0,\qquad \left[\hat{H},\hat{H}\right]=0,
\end{equation}
implying that, from the point of view of the Heisenberg equation of motion in the Heisenberg picture of time evolution,
it is a conserved quantity, $\hat{H}(t)=\hat{H}(t=0)=\hat{H}$, without any explicit time dependency.

Within the Schr\"odinger picture for time evolution, the Noether charges for spatial translations and Galilean boosts are given as,
\begin{equation}
\hat{\vec{T}}(t)=\hat{\vec{p}}_0\,-\,t\,\vec{F}\,\mathbb{I},\quad
\hat{\vec{\gamma}}(t)=-m\hat{\vec{x}}_0 + t\,\hat{\vec{p}}_0 - \frac{1}{2}t^2\,\vec{F}\,\mathbb{I},\qquad
\hat{\vec{T}}(t)^{\,\dagger}=\hat{\vec{T}}(t),\quad
\hat{\vec{\gamma}}(t)^{\,\dagger}=\hat{\vec{\gamma}}(t).
\end{equation}
Note that $\hat{\vec{T}}(t=0)=\hat{\vec{p}}_0$ and $\hat{\vec{\gamma}}(t=0)=-m\hat{\vec{x}}_0$.
These dynamical conserved quantum observables are such that,
\begin{equation}
\left[\hat{\vec{T}}(t),\hat{H}\right]=i\hbar\,\vec{F}\,\mathbb{I},\qquad
\left[\hat{\vec{\gamma}}(t),\hat{H}\right]=-i\hbar\,\hat{\vec{T}}(t),
\end{equation}
while their algebra is given as,
\begin{equation}
\left[\hat{T}_i(t),\hat{T}_j(t)\right]=0,\qquad
\left[\hat{T}_i(t),\hat{\gamma}_i(t)\right]=i\hbar\,m\,\delta_{ij}\,\mathbb{I},\qquad
\left[\hat{\gamma}_i(t),\hat{\gamma}_j(t)\right]=0.
\end{equation}

\subsection{Time evolution in the Heisenberg picture}
\label{Sect5.3}

Within the Heisenberg picture the time evolution of a quantum observable $\hat{A}_H(t)$ is governed by the Heisenberg equation of motion
(which is direct correspondence with its classical Hamiltonian equation of motion),
\begin{equation}
i\hbar\frac{d\hat{A}_H(t)}{dt}=i\hbar\frac{\partial\hat{A}_H(t)}{\partial t} + \left[\hat{A}_H(t),\hat{H}\right],
\end{equation}
in which the time independency of the Hamiltonian operator within the Heisenberg picture is already accounted for (namely $\hat{H}_H(t)=\hat{H}$),
while the partial time derivative term on the r.h.s.~of this relation only acts on the possible explicit time dependency
of the observable $\hat{A}(t)$ in the Schr\"odinger picture. If $\hat{A}(t)$ denotes that observable in the Schr\"odinger
picture as obtained through quantisation at $t=0$ of the classical dynamics as discussed in Sect.\ref{Sect5.2}, the general solution
for the time evolution of the observable in the Heisenberg picture is thus given by the following unitary transformation
involving the generator for finite translations in time,
\begin{equation}
\hat{A}_H(t)=e^{\frac{i}{\hbar}t\,\hat{H}}\,\hat{A}(t)\,e^{-\frac{i}{\hbar}t\,\hat{H}}.
\end{equation}

In particular by applying the relevant Baker-Campbell-Hausdorff formula (for $e^A B e^{-A}$),
one readily finds for the time evolution of the basic phase space observables in the Heisenberg picture,
\begin{equation}
\hat{\vec{x}}_H(t)=\hat{\vec{x}}_0+\frac{t}{m}\hat{\vec{p}}_0+\frac{t^2}{2m}\vec{F}\,\mathbb{I},\qquad
\hat{\vec{p}}_H(t)=\hat{\vec{p}}_0 + t\,\vec{F}\,\mathbb{I},
\end{equation}
in direct correspondence with their classical solutions (on account of the linearity of the classical and quantum
equations of motion of the system). Similarly one finds for the Noether charges in the Heisenberg picture,
\begin{equation}
\hat{\vec{T}}_H(t)=\hat{\vec{p}}_H(t)-t\,\vec{F}\,\mathbb{I}=\hat{\vec{p}}_0,\qquad
\hat{\vec{\gamma}}_H(t)=-m\hat{\vec{x}}_H(t)+t\,\hat{\vec{p}}_H(t)-\frac{1}{2}t^2\vec{F}\,\mathbb{I}=-m\hat{\vec{x}}_0,
\end{equation}
establishing that these are indeed constants of the motion, in spite of the fact that they do not commute with the Hamiltonian,
with $\left[\hat{\vec{T}}_H(t),\hat{H}\right]=i\hbar\vec{F}\,\mathbb{I}$ and $\left[\hat{\vec{\gamma}}_H(t),\hat{H}\right]=-i\hbar\,\hat{\vec{T}}_H(t)$,
while they determine a Heisenberg algebra with $\left[\hat{T}_{H,i}(t),\hat{\gamma}_{H,j}(t)/m\right]=i\hbar\delta_{ij}\,\mathbb{I}$.

In a likewise manner finite symmetry transformations of quantum observables in the Heisenberg picture generated by these Noether charges
are represented by the following unitary transformations,
\begin{equation}
\hat{A}_H(t|\vec{a},\vec{v})=e^{\frac{i}{\hbar}\left(\vec{a}\cdot\hat{\vec{T}}_H(t)+\vec{v}\cdot\hat{\vec{\gamma}}_H(t)\right)}\,\hat{A}_H(t)\,
e^{-\frac{i}{\hbar}\left(\vec{a}\cdot\hat{\vec{T}}_H(t)+\vec{v}\cdot\hat{\vec{\gamma}}_H(t)\right)}.
\end{equation}
In particular one then finds,
\begin{equation}
\hat{\vec{x}}_H(t|\vec{a},\vec{v})=\hat{\vec{x}}_H(t)+\left(\vec{a}+t\,\vec{v}\right)\mathbb{I},\qquad
\hat{\vec{p}}_H(t|\vec{a},\vec{v})=\hat{\vec{p}}_H(t)+m\vec{v}\,\mathbb{I},
\end{equation}
as well as,
\begin{equation}
\hat{H}(\vec{a},\vec{v}\,)=\hat{H}-\vec{a}\cdot\vec{F}\,\mathbb{I}+\vec{v}\cdot\hat{\vec{T}}_H(t)+\frac{1}{2}m\vec{v}\,^2\,\mathbb{I}
=\frac{1}{2m}\left(\hat{\vec{p}}_0+m\vec{v}\,\mathbb{I}\right)^2 - \left(\hat{\vec{x}}_0+\vec{a}\,\mathbb{I}\right)\cdot\vec{F},
\end{equation}
indeed as should be expected given the corresponding expressions for the classical dynamics.

To conclude, let us point out that the full dynamical symmetry Lie algebra of constants of motion is thus spanned by the enveloping algebra
of the Noether charges $(\hat{\vec{T}}_H(t),\hat{\vec{\gamma}}_H(t))$ as primary generators which, as dynamical conserved charges, also
determine a generating algebra for the entire energy spectrum of the system, to which we now turn.

\subsection{Time evolution in the Schr\"odinger picture}
\label{Sect5.4}

Within the Schr\"odinger picture the time evolution of a quantum state $|\psi,t\rangle$ is governed by the Schr\"odinger equation,
\begin{equation}
i\hbar\frac{d|\psi,t\rangle}{dt}=\hat{H}\,|\psi,t\rangle,
\end{equation}
while quantum observables $\hat{A}(t)$ evolve in time only provided they possess an explicit time dependency (which is thus not generated by any
action of the Hamiltonian operator), as is the case for instance for the Noether charges $(\hat{\vec{T}}(t),\hat{\vec{\gamma}}(t))$ in the Schr\"odinger
picture.

The space of quantum states provides a representation of the Heisenberg algebra spanned by the basic phase space operators
$(\hat{\vec{x}}_0,\hat{\vec{p}}_0)$, leading to the well known configuration and momentum space wave function representations
of that algebra using normalised position and momentum eigenstates such that,
\begin{equation}
\hat{\vec{x}}_0|\vec{x}\,\rangle=\vec{x}\,|\vec{x}\,\rangle,\quad
\hat{\vec{p}}_0|\vec{p}\,\rangle=\vec{p}\,|\vec{p}\,\rangle,\quad
\langle\vec{x}_1|\vec{x}_2\rangle=\delta^{(d)}(\vec{x}_1-\vec{x}_2),\quad
\langle\vec{p}_1|\vec{p}_2\rangle=\delta^{(d)}(\vec{p}_1-\vec{p}_2).
\end{equation}
Consequently,
\begin{equation}
\int_{(\infty)}d^d\vec{x}\,|\vec{x}\,\rangle\,\langle\vec{x}\,|=\mathbb{I}=\int_{(\infty)}d^d\vec{p}\,|\vec{p}\,\rangle\,\langle\vec{p}\,|,\qquad
|\varphi\rangle=\int_{(\infty)}d^d\vec{x}\,|\vec{x}\,\rangle\,\varphi(\vec{x})=\int_{(\infty)}d^d\vec{p}\,|\vec{p}\,\rangle\,\tilde{\varphi}(\vec{p}\,),
\end{equation}
with the configuration and momentum space wave functions representing the abstract state $|\varphi\rangle$ defined by
$\varphi(\vec{x}\,)=\langle\vec{x}\,|\varphi\rangle$ and $\tilde{\varphi}(\vec{p}\,)=\langle\vec{p}\,|\varphi\rangle$, respectively. Finally the local phase
of the position and momentum eigenstates may be chosen such that the operators $(\hat{\vec{x}}_0,\hat{\vec{p}}_0)$ possess the following
wave function representations\cite{Victor2},
\begin{equation}
\langle\vec{x}\,|\hat{\vec{x}}_0|\varphi\rangle=\vec{x}\,\varphi(\vec{x}\,),\quad
\langle\vec{x}\,|\hat{\vec{p}}_0|\varphi\rangle=-i\hbar\vec{\nabla}_{\vec{x}}\,\varphi(\vec{x}\,),\quad
\langle\vec{p}\,|\hat{\vec{x}}_0|\varphi\rangle=i\hbar\vec{\nabla}_{\vec{p}}\,\tilde{\varphi}(\vec{p}\,),\quad
\langle\vec{p}\,|\hat{\vec{p}}_0|\varphi\rangle=\vec{p}\,\tilde{\varphi}(\vec{p}\,),\quad
\end{equation}
which in particular implies the following change of basis overlaps and Fourier transformations relating the wave functions $\varphi(\vec{x}\,)$
and $\tilde{\varphi}(\vec{p}\,)$, in which a final relative global phase choice is implicit,
\begin{equation}
\langle\vec{x}\,|\vec{p}\,\rangle=\frac{1}{(2\pi\hbar)^{d/2}}\,e^{\frac{i}{\hbar}\vec{x}\cdot\vec{p}},\qquad
\langle\vec{p}\,|\vec{x}\,\rangle=\frac{1}{(2\pi\hbar)^{d/2}}\,e^{-\frac{i}{\hbar}\vec{x}\cdot\vec{p}}.
\end{equation}
Consequently the Noether charges have the following wave function representations, namely in configuration space,
\begin{equation}
\hat{\vec{T}}(t)\ :\quad -i\hbar\vec{\nabla}_{\vec{x}}\,-\,t\,\vec{F}\ ;\qquad
\hat{\vec{\gamma}}(t)\ :\quad -m\vec{x}\,-\,i\hbar\,t\,\vec{\nabla}_{\vec{x}}\,-\,\frac{1}{2}t^2\,\vec{F},
\end{equation}
and in momentum space,
\begin{equation}
\hat{\vec{T}}(t)\ :\quad \vec{p}\,-\,t\,\vec{F}\ ;\qquad
\hat{\vec{\gamma}}(t)\ :\quad -i\hbar\,m\,\vec{\nabla}_{\vec{p}}\,+\,t\,\vec{p}\,-\,\frac{1}{2}t^2\,\vec{F}.
\end{equation}

However since the Noether charges themselves define what is essentially another Heisenberg algebra, so that $\hat{\vec{T}}(t)$ on the one hand,
and $\hat{\vec{\gamma}}(t)$ on the other hand, determine a complete set of commuting observables, one may consider in a likewise manner
two further bases in Hilbert space spanned by the normalised eigenstates of each of these Noether charges, with time independent eigenvalues
since these are dynamical conserved observables. Hence let us consider the following eigenstates,
\begin{equation}
\hat{\vec{T}}(t)\,|\vec{T},t\rangle=\vec{T}\,|\vec{T},t\rangle,\qquad
\hat{\vec{\gamma}}(t)\,|\vec{\gamma},t\rangle = \vec{\gamma}\,|\vec{\gamma},t\rangle,
\end{equation}
normalised such that
\begin{equation}
\langle\vec{T}_1,t|\vec{T}_2,t\rangle=\delta^{(d)}(\vec{T}_1-\vec{T}_2),\qquad
\langle\gamma_1,t|\vec{\gamma}_2,t\rangle = \delta^{(d)}(\vec{\gamma}_1-\gamma_2).
\end{equation}
Consequently
\begin{equation}
\int_{(\infty)}d^d\vec{T}\,|\vec{T},t\rangle\,\langle\vec{T},t|=\mathbb{I}=\int_{(\infty)}d^d\vec{\gamma}\,|\vec{\gamma},t\rangle\,\langle\vec{\gamma},t|,
\end{equation}
thereby leading to the $\vec{T}$- and $\vec{\gamma}$-wave function representations of quantum states,
\begin{equation}
|\varphi\rangle=\int_{(\infty)}d^d\vec{T}\,|\vec{T},t\rangle\,\varphi_{\vec{T}}(\vec{T},t)
=\int_{(\infty)}d^d\vec{\gamma}\,|\vec{\gamma},t\rangle\,\varphi_{\vec{\gamma}}(\vec{\gamma},t),
\end{equation}
where
\begin{equation}
\varphi_{\vec{T}}(\vec{T},t)=\langle\vec{T},t|\varphi\rangle,\qquad
\varphi_{\vec{\gamma}}(\vec{\gamma},t)=\langle\vec{\gamma},t|\varphi\rangle.
\end{equation}
Finally the local phase of the $\hat{\vec{T}}$- and $\hat{\vec{\gamma}}$-eigenstates may be chosen in such a way that the Noether charges have the following
$\vec{T}$- and $\vec{\gamma}$-wave function representations, namely in the $\vec{T}$-representation,
\begin{equation}
\hat{\vec{T}}(t)\ :\quad \vec{T}\ ;\qquad
\hat{\vec{\gamma}}(t)\ :\quad -i\hbar\,m\,\vec{\nabla}_{\vec{T}},
\end{equation}
and in the $\vec{\gamma}$-representation,
\begin{equation}
\hat{\vec{T}}(t)\ :\quad i\hbar\,m\,\vec{\nabla}_{\vec{\gamma}}\ ;\qquad
\hat{\vec{\gamma}}(t)\ :\quad \vec{\gamma} .
\end{equation}
Quite obviously one then has as well, up to some phase factor $\Phi(\vec{T},\vec{\gamma},t)$ still to be specified or identified at least implicitly (as will be done
hereafter),
\begin{equation}
\langle\vec{T},t|\vec{\gamma},t\rangle=\frac{1}{(2\pi m\hbar )^{d/2}}\,e^{\frac{i}{m\hbar}\vec{T}\cdot\vec{\gamma}}\,e^{i\Phi(\vec{T},\vec{\gamma},t)},\quad
\langle\vec{\gamma},t|\vec{T},t\rangle=\frac{1}{(2\pi m\hbar )^{d/2}}\,e^{-\frac{i}{m\hbar}\vec{T}\cdot\vec{\gamma}}\,e^{-i\Phi(\vec{T},\vec{\gamma},t)}.
\label{eq:Tg-overlap}
\end{equation}

\subsubsection{Energy eigenstates}
\label{Sect5.4.1}

Before determining the $\vec{x}$- and $\vec{p}$-representations of the $\hat{\vec{T}}$- and $\hat{\vec{\gamma}}$-eigenstates, let us turn to the identification
of the energy eigenstates of the system, in the form
\begin{equation}
|\psi_E,t\rangle=e^{-\frac{i}{\hbar}t\,E}\,|E\rangle,\qquad
\hat{H}|E\rangle = E\,|E\rangle.
\end{equation}
Given that the potential $V(\vec{x}\,)=-\vec{x}\cdot\vec{F}$ is unbounded below (and above), the system does not possess a ground state, a feature which
in itself is not a physical inconsistency. Provided it remains decoupled from any external interaction other than the constant force field,
the system displays stability under time evolution in any of its possible energy configurations, of which the continuum spectrum covers the entire
real line, $-\infty<E<+\infty$. Consequently energy eigenstates are to be normalised according to
\begin{equation}
\langle E_1|E_2\rangle=\delta(E_1-E_2),\qquad
E_1,E_2\in\mathbb{R}.
\end{equation}
Note well however that energy eigen-levels will (for $d\ge 3$) possess a degeneracy related to the SO($d-1$) symmetry of the dynamics under
rotations in all those planes perpendicular to the constant force vector $\vec{F}$.

For later convenience let us introduce the unit vector $\hat{F}$ in the direction of $\vec{F}$,
\begin{equation}
\hat{F}=\frac{1}{F}\vec{F},\qquad
F=|\vec{F}|>0,\qquad
\vec{F}=F\,\hat{F}.
\end{equation}
Any vector, say $\vec{x}$ of $\vec{p}$, is then to be decomposed as follows in terms of components parallel and perpendicular relative to $\hat{F}$,
\begin{equation}
\vec{x}=x_\parallel\,\hat{F}\,+\,\vec{x}_\bot,\quad
\vec{p}=p_\parallel\,\hat{F}\,+\,\vec{p}_\bot,\qquad
x_\parallel=\vec{x}\cdot\hat{F},\quad
p_\parallel=\vec{p}\cdot\hat{F},\qquad
\vec{x}_\bot\cdot\hat{F}=0=\vec{p}_\bot\cdot\hat{F}.
\end{equation}

In order to solve the stationary Schr\"odinger equation it is best to work in the $\vec{p}$-repre\-sen\-ta\-tion, leading to,
\begin{equation}
\left(\frac{1}{2m}\vec{p}\,^2-i\hbar\,\vec{F}\cdot\vec{\nabla}_{\vec{p}}\right)\,\tilde{\varphi}_E(\vec{p}\,) = E\,\tilde{\varphi}_E(\vec{p}\,),\qquad
\tilde{\varphi}_E(\vec{p}\,)=\langle\vec{p}\,|E\rangle,
\end{equation}
or equivalently,
\begin{equation}
-i\hbar\,F\,\frac{\partial}{\partial p_\parallel}\,\tilde{\varphi}_E(p_\parallel,\vec{p}_\bot)
=\left(E-\frac{1}{2m}\vec{p}_\bot^{\,2}-\frac{1}{2m}p_\parallel^2\right)\,\tilde{\varphi}_E(p_\parallel,\vec{p}_\bot).
\end{equation}
The general solution, properly normalised, is then of the form,
\begin{equation}
\tilde{\varphi}_E(\vec{p}\,)=\langle\vec{p}\,| E\rangle = \frac{1}{\sqrt{2\pi\hbar F}}\,\tilde{\varphi}_\bot(\vec{p}_\bot)\,{\rm exp}\left\{\frac{i}{\hbar F}
\left[\left(E-\frac{1}{2m}\vec{p}_\bot^{\,2}\right)\,p_\parallel\,-\,\frac{1}{6m}\,p^3_\parallel\right]\right\},\quad E\in\mathbb{R},
\end{equation}
where $\tilde{\varphi}_\bot(\vec{p}_\bot)$ is some arbitrarily chosen function parametrising the degeneracy of the energy levels under SO($d-1$) rotations
and normalised such that
\begin{equation}
\int_{(\infty)}d^{d-1}\vec{p}_\bot\,|\tilde{\varphi}_\bot(\vec{p}_\bot)|^2=1.
\end{equation}
If need be, a basis for these functions $\tilde{\varphi}_\bot(\vec{p}_\bot)$ may be constructed in terms of Bessel functions and SO($d-1$) hyperspherical harmonics
in the space $\vec{p}_\bot$.

Through inverse Fourier transformations the $\vec{x}$-representation of these states is obtained from,
\begin{eqnarray}
&&\varphi_E(x_\parallel,\vec{x}_\bot) = \varphi_E(\vec{x}\,) =
\langle\vec{x}\,|E\rangle=\int_{(\infty)}\frac{d^d\vec{p}}{(2\pi\hbar)^{d/2}}\,e^{\frac{i}{\hbar}\vec{x}\cdot\vec{p}}\,\tilde{\varphi}_E(\vec{p}\,) = \\
&=& \frac{1}{\sqrt{2\pi\hbar F}}\int_{-\infty}^\infty \frac{dp_\parallel}{\sqrt{2\pi\hbar}}\,e^{\frac{i}{\hbar}x_\parallel p_\parallel}\,
e^{\frac{i}{\hbar F}\left[\left(E-\frac{1}{2m}\vec{p}_\bot^{\,2}\right)\,p_\parallel\,-\,\frac{1}{6m}\,p^3_\parallel\right]}
\times\,
\int_{(\infty)}\frac{d^{d-1}\vec{p}_\bot}{(2\pi\hbar)^{(d-1)/2}}\,e^{\frac{i}{\hbar}\vec{x}_\bot\cdot\vec{p}_\bot}\,\tilde{\varphi}_\bot(\vec{p}_\bot), \nonumber
\end{eqnarray}
and this in particular in terms of the Airy function of the first kind for what the dependency on the cartesian coordinate $x_\parallel$ aligned with the constant
force field is concerned\cite{Victor} (which results from the integral over $p_\parallel$).

\subsubsection{$\hat{\vec{T}}(t)$-eigenstates}
\label{Sect5.4.2}

Working once again in the $\vec{p}$-representation, $\hat{\vec{T}}(t)$-eigenstates $|\vec{T},t\rangle$ are characterised by the following eigenvalue equation,
\begin{equation}
\left(\vec{p}\,-\,t\,\vec{F}\right)\,\tilde{\varphi}_{\vec{T}}(\vec{p},t)=\vec{T}\,\tilde{\varphi}_{\vec{T}}(\vec{p},t),\qquad
\tilde{\varphi}_{\vec{T}}(\vec{p},t)=\langle\vec{p}\,|\vec{T},t\rangle.
\end{equation}
Therefore necessarily
\begin{equation}
\tilde{\varphi}_{\vec{T}}(\vec{p},t)=\delta^{(d)}\left(\vec{p}-t\,\vec{F} - \vec{T}\right)\,\tilde{\chi}_{\vec{T}}(t),
\end{equation}
where $\tilde{\chi}_{\vec{T}}(t)$ is a still to be determined function of $t$ only, such that the states $|\vec{T},t\rangle$ obey the Schr\"odinger equation,
namely,
\begin{equation}
\left(\frac{1}{2m}\vec{p}^{\,2}-i\hbar\vec{F}\cdot\vec{\nabla}_{\vec{p}}\right)\,\tilde{\varphi}_{\vec{T}}(\vec{p},t)=
i\hbar\frac{\partial}{\partial t}\,\tilde{\varphi}_{\vec{T}}(\vec{p},t).
\end{equation}
As a result one must have,
\begin{equation}
i\hbar\,\frac{d}{dt}\tilde{\chi}_{\vec{T}}(t)=\frac{1}{2m}\left(\vec{T}_\bot^2+(T_\parallel + t F)^2\right)\,\tilde{\chi}_{\vec{T}}(t)
=\frac{1}{2m}\left(\vec{T}^2+2t\,\vec{T}\cdot\vec{F}+t^2\vec{F}^2\right)\,\tilde{\chi}_{\vec{T}}(t).
\end{equation}

By choosing appropriately the hitherto unspecified local phase of the states $|\vec{T},t\rangle$, and imposing the normalisation of the states $|\vec{T},t\rangle$
as indicated above, one thus finds,
\begin{equation}
\langle\vec{p}\,|\vec{T},t\rangle=\tilde{\varphi}_{\vec{T}}(\vec{p},t)=\delta^{(d)}\left(\vec{p}-t\vec{F}-\vec{T}\right)\,
e^{-\frac{i}{\hbar}\frac{t}{2m}\left(\vec{T}^2+t\,\vec{T}\cdot\vec{F}+\frac{t^2}{3}\,\vec{F}^2\right)},
\end{equation}
as well as in the $\vec{x}$-representation,
\begin{equation}
\langle\vec{x}\,|\vec{T},t\rangle=\frac{1}{(2\pi\hbar)^{d/2}}\,e^{\frac{i}{\hbar}\vec{x}\cdot(\vec{T}+t\vec{F})}\,
e^{-\frac{i}{\hbar}\frac{t}{2m}\left(\vec{T}^2+t\,\vec{T}\cdot\vec{F}+\frac{t^2}{3}\,\vec{F}^2\right)}.
\end{equation}

Incidentally note that as a matter of fact the $\hat{\vec{T}}(t)$-eigenstates coincide essentially with the momentum eigenstates but with a time dependency
in their momentum value and in their phase factor, in the form,
\begin{equation}
|\vec{T},t\rangle=e^{-\frac{i}{\hbar}\frac{t}{2m}\left(\vec{T}^2+t\,\vec{T}\cdot\vec{F}+\frac{t^2}{3}\,\vec{F}^2\right)}\,|\vec{p}=\vec{T}+t\vec{F}\rangle.
\end{equation}

\subsubsection{$\hat{\vec{\gamma}}(t)$-eigenstates}
\label{Sect5.4.3}

Working once again in the $\vec{p}$-representation, $\hat{\vec{\gamma}}(t)$-eigenstates $|\vec{\gamma},t\rangle$ are characterised by the following
eigenvalue equation,
\begin{equation}
\left(-i\hbar\,m\,\vec{\nabla}_{\vec{p}} + t\,\vec{p} -\frac{1}{2}t^2\,\vec{F}\right)\,\tilde{\varphi}_{\vec{\gamma}}(\vec{p},t)
=\vec{\gamma}\,\tilde{\varphi}_{\vec{\gamma}}(\vec{p},t),\qquad
\tilde{\varphi}_{\vec{\gamma}}(\vec{p},t)=\langle\vec{p}\,|\vec{\gamma},t\rangle.
\end{equation}
Therefore necessarily
\begin{equation}
\tilde{\varphi}_{\vec{\gamma}}(\vec{p},t)=e^{\frac{i}{\hbar m}\left(\vec{\gamma}\cdot\vec{p}+\frac{t^2}{2}\vec{F}\cdot\vec{p}-\frac{t}{2}\vec{p}^{\,2}\right)}\,
\tilde{\chi}_{\vec{\gamma}}(t),
\end{equation}
where $\tilde{\chi}_{\vec{\gamma}}(t)$ is a still to be determined function of $t$ only, such that the states $|\vec{\gamma},t\rangle$ obey the Schr\"odinger equation,
namely,
\begin{equation}
\left(\frac{1}{2m}\vec{p}^{\,2}-i\hbar\vec{F}\cdot\vec{\nabla}_{\vec{p}}\right)\,\tilde{\varphi}_{\vec{\gamma}}(\vec{p},t)=
i\hbar\frac{\partial}{\partial t}\,\tilde{\varphi}_{\vec{\gamma}}(\vec{p},t).
\end{equation}
As a result one must have,
\begin{equation}
i\hbar\,\frac{d}{dt}\tilde{\chi}_{\vec{\gamma}}(t)=\frac{1}{m}\left(\vec{F}\cdot\vec{\gamma}+\frac{t^2}{2}\vec{F}^2\right)\,\tilde{\chi}_{\vec{\gamma}}(t).
\end{equation}

By choosing appropriately the hitherto unspecified local phase of the states $|\vec{\gamma},t\rangle$, and imposing the normalisation of the states
$|\vec{\gamma},t\rangle$ as indicated above, one thus finds,
\begin{equation}
\langle\vec{p}\,|\vec{\gamma},t\rangle=\tilde{\varphi}_{\vec{\gamma}}(\vec{p},t)=\frac{1}{(2\pi\hbar\, m)^{d/2}}\,
e^{-\frac{i}{\hbar}\frac{t}{m}\left(\vec{F}\cdot\vec{\gamma}+\frac{t^2}{6}\vec{F}^2\right)}\,
e^{\frac{i}{\hbar m}\left(\vec{\gamma}\cdot\vec{p}+\frac{t^2}{2}\vec{F}\cdot\vec{p}-\frac{t}{2}\vec{p}^{\,2}\right)},
\end{equation}
as well as in the $\vec{x}$-representation,
\begin{equation}
\langle\vec{x}\,|\vec{\gamma},t\rangle=\frac{1}{(2\pi\hbar\,it)^{d/2}}\,e^{\frac{i}{\hbar}\frac{1}{2mt}\left(m\vec{x}+\vec{\gamma}+\frac{t^2}{2}\vec{F}\right)^2}\,
e^{-\frac{i}{\hbar}\frac{t}{m}\left(\vec{F}\cdot\vec{\gamma}+\frac{t^2}{6}\vec{F}^2\right)}.
\end{equation}

The knowledge of the configuration and momentum wave function representations of the $\hat{\vec{T}}(t)$- and $\hat{\vec{\gamma}}(t)$-eigenstates,
enables one to also compute their overlap, in view of the earlier remark that these operators essentially define a Heisenberg algebra.
Completing the relevant momentum integrations for the product $\langle\vec{T},t|\vec{p}\,\rangle\langle\vec{p}\,|\vec{\gamma},t\rangle$, one finds,
\begin{equation}
\langle\vec{T},t|\vec{\gamma},t\rangle=\frac{1}{(2\pi\hbar\,m)^{d/2}}\,
e^{\frac{i}{\hbar}\frac{1}{m}\vec{T}\cdot\vec{\gamma}}\,e^{\frac{i}{\hbar}\frac{t}{m}\left(\vec{T}+t\vec{F}\right)^2} ,
\end{equation}
a result which is certainly perfectly in line with the Heisenberg nature of the algebra defined by these two Noether charges,
and finally determines the phase factor introduced in (\ref{eq:Tg-overlap}) as being given by  $\Phi(\vec{T},\vec{\gamma},t)=t(\vec{T}+t\vec{F})^2/(\hbar m)$.

\subsubsection{Finite quantum spatial translations and Galilean boosts}
\label{Sect5.4.4}

Given the Noether charges in the Schr\"odinger picture expressed as
\begin{equation}
\hat{\vec{T}}(t)=\hat{\vec{p}}_0-t\vec{F}\,\mathbb{I},\qquad
\hat{\vec{\gamma}}(t)=-m\hat{\vec{x}}_0+t\hat{\vec{p}}_0-\frac{1}{2}t^2\vec{F}\,\mathbb{I},
\end{equation}
and using the different wave function representations constructed above, it is a simple matter to check explicitly that indeed
the states $|\vec{T},t\rangle$ and $|\vec{\gamma},t\rangle$ are eigenstates of these operators with time independent and thus
conserved eigenvalues, which obey as well the time dependent Schr\"odinger equation, namely,
\begin{equation}
\hat{\vec{T}}(t)|\vec{T},t\rangle=\vec{T}\,|\vec{T},t\rangle,\quad
\hat{\vec{\gamma}}(t)|\vec{\gamma},t\rangle=\vec{\gamma}\,|\vec{\gamma},t\rangle,\quad
\hat{H}|\vec{T},t\rangle=i\hbar\frac{d}{d t}|\vec{T},t\rangle,\quad
\hat{H}|\vec{\gamma},t\rangle=i\hbar\frac{d}{d t}|\vec{\gamma},t\rangle.
\end{equation}
Even though dynamical with an explicit time dependency within the Schr\"odinger picture, these Noether charges are nevertheless constants of the motion,
in full agreement with Noether's first theorem within this quantum context.

Let us now consider finite transformations generated by these charges, within the Schr\"o\-din\-ger picture, beginning with spatial translations.
Acting on the basic phase space observables, a straightforward calculation finds,
\begin{equation}
e^{\frac{i}{\hbar}\vec{a}\cdot\hat{\vec{T}}(t)}\,\hat{\vec{x}}_0\,e^{-\frac{i}{\hbar}\vec{a}\cdot\hat{\vec{T}}(t)} = \hat{\vec{x}}_0\,+\,\vec{a}\,\mathbb{I},\qquad
e^{\frac{i}{\hbar}\vec{a}\cdot\hat{\vec{T}}(t)}\,\hat{\vec{p}}_0\,e^{-\frac{i}{\hbar}\vec{a}\cdot\hat{\vec{T}}(t)} = \hat{\vec{p}}_0,
\end{equation}
as it should be. Given that the $\hat{\vec{T}}(t)$ eigenstates are essentially momentum eigenstates with a time dependency in their momentum eigenvalue
and in their phase, it readily follows that finite spatial translations transform momentum eigenstates according to,
\begin{equation}
e^{-\frac{i}{\hbar}\vec{a}\cdot\hat{\vec{T}}(t)}\,|\vec{p}\,\rangle=  e^{-\frac{i}{\hbar}\,\vec{a}\cdot(\vec{p}-t\vec{F})}\,|\vec{p}\,\rangle.
\end{equation}
Given the change of basis overlaps $\langle\vec{p}\,|\vec{x}\,\rangle$, from this result one then establishes as well that,
\begin{equation}
e^{-\frac{i}{\hbar}\vec{a}\cdot\hat{\vec{T}}(t)}\,|\vec{x}\,\rangle=e^{\frac{i}{\hbar}\,t\,\vec{a}\cdot\vec{F}}\,|\vec{x}+\vec{a}\,\rangle.
\end{equation}

Considering now finite quantum Galilean boost transformations, one finds, thus within the Schr\"odinger picture,
\begin{equation}
e^{\frac{i}{\hbar}\vec{v}\cdot\hat{\vec{\gamma}}(t)}\,\hat{\vec{x}}_0\,e^{-\frac{i}{\hbar}\vec{v}\cdot\hat{\vec{\gamma}}(t)} = \hat{\vec{x}}_0\,+\,t\,\vec{v}\,\mathbb{I},\qquad
e^{\frac{i}{\hbar}\vec{v}\cdot\hat{\vec{\gamma}}(t)}\,\hat{\vec{p}}_0\,e^{-\frac{i}{\hbar}\vec{v}\cdot\hat{\vec{\gamma}}(t)} = \hat{\vec{p}}_0\,+\,m\vec{v}\,\mathbb{I},
\end{equation}
as it should be. Furthermore using the change of basis $\langle\vec{\gamma},t|\vec{p}\,\rangle$, an explicit calculation establishes that under Galilean boosts
momentum eigenstates transform according to,
\begin{equation}
e^{-\frac{i}{\hbar}\vec{v}\cdot\hat{\vec{\gamma}}(t)}\,|\vec{p}\,\rangle=e^{\frac{i}{\hbar}\left(\frac{t^2}{2}\vec{v}\cdot\vec{F}-t\,\vec{v}\cdot\vec{p}
-\frac{t}{2}m\vec{v}^{\,2}\right)}\,|\vec{p}+m\vec{v}\,\rangle,
\end{equation}
from which it follows as well that, or else using the overlaps $\langle\vec{\gamma},t|\vec{x}\,\rangle$,
\begin{equation}
e^{-\frac{i}{\hbar}\vec{v}\cdot\hat{\vec{\gamma}}(t)}\,|\vec{x}\,\rangle=
e^{\frac{i}{\hbar}\left(m\vec{v}\cdot\vec{x}+\frac{1}{2}mt\,\vec{v}^{\,2}+\frac{t^2}{2}\vec{v}\cdot\vec{F}\right)}\,|\vec{x}+t\,\vec{v}\,\rangle.
\end{equation}

Finally let us address the action of these symmetry transformations on the energy eigenstates. First it is straightforward to establish that one has,
within the Schr\"odinger picture,
\begin{equation}
e^{\frac{i}{\hbar}\vec{a}\cdot\hat{\vec{T}}(t)}\,\hat{H}\,e^{-\frac{i}{\hbar}\vec{a}\cdot\hat{\vec{T}}(t)}
=  \hat{H}-\vec{a}\cdot\vec{F}\,\mathbb{I},\qquad
e^{\frac{i}{\hbar}\vec{v}\cdot\hat{\vec{\gamma}}(t)}\,\hat{H}\,e^{-\frac{i}{\hbar}\vec{v}\cdot\hat{\vec{\gamma}}(t)}
=\hat{H}+\vec{v}\cdot\hat{\vec{T}}(t)+\frac{1}{2}m\vec{v}^{\,2}\,\mathbb{I},
\end{equation}
indeed as ought to be the case in view of the corresponding results for the classical dynamics. Consider now an arbitrary energy eigenstate,
\begin{equation}
|\psi_E,t\rangle=e^{-\frac{i}{\hbar}t\,E}\,|E\rangle,\qquad
\hat{H}|E\rangle=E\,|E\rangle,\qquad
\hat{H}|\psi_E,t\rangle=E\,|\psi_E,t\rangle=i\hbar\frac{d}{dt}|\psi_E,t\rangle.
\end{equation}
Within the Schr\"odinger picture and under a spatial translation this state is transformed into
\begin{equation}
|\psi_E,t;\vec{a}\,\rangle=e^{-\frac{i}{\hbar}\vec{a}\cdot\hat{\vec{T}}(t)}\,|\psi_E,t\rangle=
e^{-\frac{i}{\hbar}t\,E}\,e^{-\frac{i}{\hbar}\vec{a}\cdot\hat{\vec{T}}(t)}\,|E\rangle
=e^{-\frac{i}{\hbar}t(E-\vec{a}\cdot\vec{F})}\,e^{-\frac{i}{\hbar}\vec{a}\cdot\hat{\vec{p}}_0}\,|E\rangle.
\end{equation}
Applying the identity established above for the transformed Hamiltonian itself, it readily follows that
\begin{equation}
\hat{H}|\psi_E, t;\vec{a}\,\rangle=\left(E-\vec{a}\cdot\vec{F}\right)\,|\psi_E, t;\vec{a}\,\rangle=i\hbar\frac{d}{dt}|\psi_E,t;\vec{a}\,\rangle.
\end{equation}
The transformed state $|\psi_E,t;\vec{a}\,\rangle$ is thus some energy eigenstate of energy $(E-\vec{a}\cdot\vec{F})$ which solves the
Schr\"odinger equation. The dynamical conserved Noether charge for spatial translations, $\hat{\vec{T}}(t)$, is indeed a spectrum generating symmetry
operator mapping between energy eigenstates and between general solutions to the quantum dynamics.

Consider now the Galilean boost transformed state,
\begin{equation}
|\psi_E,t;\vec{v}\,\rangle=e^{-\frac{i}{\hbar}\vec{v}\cdot\hat{\vec{\gamma}}(t)}\,|\psi_E,t\rangle
=e^{-\frac{i}{\hbar}t\,E}\,e^{-\frac{i}{\hbar}\vec{v}\cdot\hat{\vec{\gamma}}(t)}\,|E\rangle
=e^{-\frac{i}{\hbar}t\,E}\,e^{\frac{i}{\hbar}\left(\frac{t^2}{2}\vec{v}\cdot\vec{F}+m\vec{v}\cdot\hat{\vec{x}}_0-t\,\vec{v}\cdot\hat{\vec{p}}_0\right)}\,|E\rangle.
\end{equation}
In this case one finds,
\begin{equation}
\hat{H}\,|\psi_E,t;\vec{v}\,\rangle=\left(E+\frac{1}{2}m\vec{v}^{\,2}\right)\,|\psi_E,t;\vec{v}\,\rangle\,+\,e^{-\frac{i}{\hbar}\vec{v}\cdot\hat{\vec{\gamma}}(t)}\,
\vec{v}\cdot\hat{\vec{T}}(t)\,|\psi_E,t\rangle.
\end{equation}
Given the identity,
\begin{equation}
e^{-\frac{i}{\hbar}\vec{v}\cdot\hat{\vec{\gamma}}(t)}\,\hat{\vec{T}}(t)\,e^{\frac{i}{\hbar}\vec{v}\cdot\hat{\vec{\gamma}}(t)} = \hat{\vec{T}}(t) - m \vec{v}\,\mathbb{I},
\end{equation}
one may write as well,
\begin{equation}
\hat{H}\,|\psi_E,t;\vec{v}\,\rangle=\left(E+\frac{1}{2}m\vec{v}^{\,2}\right)\,|\psi_E,t;\vec{v}\,\rangle\,+\,
\left(\vec{v}\cdot\hat{\vec{T}}(t) - m\vec{v}^{\,2}\right)\,|\psi_E,t;\vec{v}\,\rangle.
\label{eq:Transf2}
\end{equation}
On the other hand when considering the Schr\"odinger equation, one has for the transformed state,
\begin{eqnarray}
i\hbar\frac{d}{dt}|\psi_E,t;\vec{v}\,\rangle &=& e^{-\frac{i}{\hbar}\vec{v}\cdot\hat{\vec{\gamma}}(t)}\,
\left[i\hbar\frac{d}{dt}+e^{\frac{i}{\hbar}\vec{v}\cdot\hat{\vec{\gamma}}(t)}\left(i\hbar\frac{d}{dt}e^{-\frac{i}{\hbar}\vec{v}\cdot\hat{\vec{\gamma}}(t)}\right)\right]
\,|\psi_E,t\rangle \nonumber \\
 &=& E\,|\psi_E,t;\vec{v}\,\rangle + e^{-\frac{i}{\hbar}\vec{v}\cdot\hat{\vec{\gamma}}(t)}
\left[e^{\frac{i}{\hbar}\vec{v}\cdot\hat{\vec{\gamma}}(t)}\left(i\hbar\frac{d}{dt}e^{-\frac{i}{\hbar}\vec{v}\cdot\hat{\vec{\gamma}}(t)}\right)\right]\,|\psi_E,t\rangle.
\end{eqnarray}
Using the following Baker-Campbell-Hausdorff identity valid for any two operators $A$ and $B$ which commute with their commutator $[A,B]$,
\begin{equation}
e^{A+B}=e^{-\frac{1}{2}[A,B]}\,e^A\,e^B,
\end{equation}
it may be established that one has,
\begin{equation}
e^{\frac{i}{\hbar}\vec{v}\cdot\hat{\vec{\gamma}}(t)}\left(i\hbar\frac{d}{dt}e^{-\frac{i}{\hbar}\vec{v}\cdot\hat{\vec{\gamma}}(t)}\right)
=\frac{1}{2}m\vec{v}^{\,2}\,\mathbb{I} - t\,\vec{v}\cdot\vec{F}\,\mathbb{I} + \vec{v}\cdot\hat{\vec{p}}_0=\frac{1}{2}m\vec{v}^{\,2}\,\mathbb{I} + \vec{v}\cdot\hat{\vec{T}}(t).
\end{equation}
Consequently the time evolution of the transformed state is such that,
\begin{equation}
i\hbar\frac{d}{dt}|\psi_E,t;\vec{v}\,\rangle=\left(E+\frac{1}{2}m\vec{v}^{\,2}\right)\,|\psi_E,t;\vec{v}\,\rangle\,+\,
e^{-\frac{i}{\hbar}\vec{v}\cdot\hat{\vec{\gamma}}(t)}\,\vec{v}\cdot\hat{\vec{T}}(t)\,|\psi_E,t\rangle=\hat{H}\,|\psi_E,t;\vec{v}\,\rangle,
\end{equation}
namely it indeed obeys the Schr\"odinger equation.

In other words, and in contradistinction to spatial translations, under a Galilean boost an energy eigenstate is not mapped into some other well defined
energy eigenstate (that could have possessed the corresponding classical energy value, $E+m\vec{v}^{\,2}/2+\vec{v}\cdot\vec{T}$, had it not been for the
quantum non-commutativity of $\hat{\vec{\gamma}}(t)$ and $\hat{\vec{T}}(t)$ as quantum operators, while the states $|\psi_E,t;\vec{v}\,\rangle$ are
not sharp $\hat{\vec{T}}(t)$-eigenstates), but rather into a specific quantum superposition of all energy eigenstates with a certain distribution characterised
by the action of the operator $\vec{v}\cdot\hat{\vec{T}}(t)$ onto the transformed state $|\psi_E,t;\vec{v}\,\rangle$ (see (\ref{eq:Transf2})). Nevertheless
Galilean boosts do map between general solutions to the Schr\"odinger equation. Therefore these observations confirm the status of the dynamical conserved
Noether charge $\hat{\vec{\gamma}}(t)$ as being a spectrum generator symmetry operator that does indeed map between solutions to the quantum dynamics,
even though it does not map energy eigenstates into energy eigenstates but rather into quantum superpositions thereof.

That Galilean boosts map energy eigenstates into distributions of energy eigenstates rather than into another energy eigenstate of specific energy
may also be confirmed by considering the mean energy distribution for transformed energy eigenstates of values $E_1$ and $E_2$, namely,
\begin{eqnarray}
\langle\psi_{E_1},t;\vec{v}\,|\hat{H}|\psi_{E_2},t;\vec{v}\,\rangle &=&
\langle\psi_{E_1},t | \left(\hat{H}+\vec{v}\cdot\hat{\vec{T}}(t)+\frac{1}{2}m\vec{v}^{\,2}\,\mathbb{I}\right)|\psi_{E_2},t \rangle \nonumber \\
&=& \left(E_1+\frac{1}{2}m\vec{v}^{\,2}\right)\,\delta(E_1-E_2)\,+\,\langle E_1|\vec{v}\cdot\hat{\vec{T}}(t)|E_2\rangle\,e^{\frac{i}{\hbar}t(E_1-E_2)}.\qquad
\end{eqnarray}
This expression illustrates explicitly that it is the quantum Noether charge for spatial translations, $\hat{\vec{T}}(t)$, which maps energy eigenstates into different
energy eigenstates, that is responsible for the spreading in energy of the action on energy eigenstates of the quantum Noether charge for Galilean boosts,
$\hat{\vec{\gamma}}(t)$. This feature is a direct consequence of the fact that these two Noether charges do not commute and generate a Heisenberg algebra
on their own. Nevertheless their status as symmetry generators and dynamical constants of the motion remains perfectly consistent within the quantum context
as well, in spite of their explicit time dependency as operators in the Schr\"odinger picture for the quantum dynamics of this system.

\section{Conclusions}
\label{SectConclusions}

Symmetries of a dynamical system are transformations of the set of all its possible configurations that map any solution to its equations of motion
to some other solution, thereby leaving invariant indeed the ensemble of all these solutions. Most standard textbooks dealing with the properties
following from the existence of symmetries set the analysis within the Lagrangian variational principle and of course rely on Noether's first theorem
in the case of continuous Lie symmetries. However, they present these matters in such a manner that it leads
one to believe, and learn, that such Lie symmetries are generated by conserved Noether charges that necessarily are time independent phase space observables,
hence that these constants of the motion necessarily commute with the Hamiltonian, and that necessarily symmetries must leave the Lagrange function
exactly invariant. However, this need not be so. Such statements are too restrictive and specific; neither do they do full justice to the true and complete
content of Noether's first theorem, nor to the full reach and meaning of it.

In complete generality however, and as a matter of fact, there do exist Lie symmetries generated by conserved Noether charges that are dynamical, namely
which possess an explicit time dependency as observables over phase space and thus do not commute with the Hamiltonian and yet define constants of the motion,
and potentially define spectrum generating Lie algebras of observables; and there exist as well symmetries that leave the system's action invariant
but only up to some total time derivative contribution. Furthermore, these features apply whether to the Lagrangian or to the Hamiltonian formulation of the dynamics,
the latter however, offering the advantage of identifying then further dynamical symmetries beyond those of the Lagrangian formulation with its symmetries
of a geometrical character in configuration space only, namely symmetry transformations that mix all phase space variables indifferently, whether
the configuration space coordinates or their canonical conjugate momenta.

In addition, within the Hamiltonian formulation, the converse statements are true
as well, namely that any dynamical constant of the motion is necessarily the Noether charge for some symmetry of the first-order Hamiltonian action
and its dynamics, which remains invariant up to a total time derivative contribution under the action of that symmetry on phase space.
All Lie symmetries are indeed so-called Noetherian symmetries, certainly within the Hamiltonian setting; there are no non-Noetherian
Lie symmetries (in spite of statements to the contrary in the literature). This remains valid upon (a complete or partial) Hamiltonian reduction (of all or some of
the canonically conjugate momenta to generalised velocities in configuration space) to a Lagrangian formulation, provided those Lie symmetry transformations
under considerations act by mixing the configuration coordinates and their generalised velocities indifferently (as they do within the Hamiltonian formulation
for these coordinates and their canonically conjugate momenta).

This less restricted and richer understanding of the possible consequences of continuous Lie symmetries in dynamical systems, directly rooted
in the complete content of Noether's first theorem, is not as widely known and integrated within textbooks as it would deserve to be,
in particular certainly when it comes to the Hamiltonian approach.
In a discussion which aims to be pedagogical as well, the first part of this work (re)visits these matters within the Hamiltonian setting in as general and streamlined
a way as possible (the sole restrictions being that the Hamiltonian does not involve any explicit time dependency, and that the system is regular, without
first- or second-class constraints, for ease of discussion). In particular it includes a detailed presentation and justification of the statement converse
to Noether's first theorem, establishing a one-to-one correspondence between Noether Lie symmetries (leaving the action invariant only up to some
total time derivative contribution) and constants of the motion.

To illustrate the general discussion in all its different aspects, three different relatively simple and straightforward classes of systems are considered as well.
Namely, first the harmonic dynamics of a collection of coupled harmonic oscillators in the basis of their normal modes, with its full algebra of dynamical
Noether constants of motion. Next, the extension of such dynamics to that of a free relativistic scalar field in Minkowski spacetime, with the identification
of all its dynamical conserved Noether charges constructed out of its time dependent observables that create and annihilate single quanta of each
of the field's oscillating modes. And finally, the dynamics of a nonrelativistic particle in euclidean space subjected to a constant force, and its symmetries
under spatial translations and Galilean boosts, with a detailed analysis specifically of its quantum dynamics and quantum conserved Noether charges
whether in the Heisenberg or in the Schr\"odinger picture for quantum time evolution, as they may be identified from these two classes of global
Lie symmetries of that system and be shown to be indeed spectrum generating operators.

One last general consideration may be worth pointing out, to again confirm the complete consistency of having dynamical quantum constants of the motion
with their explicit time dependency as conserved Noether charges which generate continuous symmetries and yet do not commute with the quantum Hamiltonian,
and now in the case when even the Hamiltonian itself possesses an explicit time dependency and does not define generally itself a conserved quantity,
namely $H(t)\equiv H(q^\alpha,p_\alpha;t)$.

Consider the quantum system to have been quantised at time $t=t_0$, with in particular (thus
 in the Schr\"odinger picture)  its quantum Hamiltonian $\hat{H}(t)$, some arbitrary quantum observable $\hat{A}(t)$, and some dynamical
 Noether charge $\hat{Q}(t)$ with its value $\hat{Q}_0=\hat{Q}(t_0)$.
In the Schr\"odinger picture time evolution is determined from the Schr\"odinger equation,
\begin{equation}
i\hbar\frac{d}{dt}|\psi,t\rangle = \hat{H}(t)\,|\psi, t\rangle ,
\end{equation}
of which the solution is given in the form,
\begin{equation}
|\psi, t\rangle = U(t,t_0)\,|\psi, t_0\rangle,
\end{equation}
where $|\psi,t_0\rangle$ is the initial quantum state at $t=t_0$, and $U(t,t_0)$ is the unitary quantum evolution operator such that
$U^\dagger(t_2,t_1)=U^{-1}(t_2,t_1)=U(t_1,t_2)$, $U(t,t)=\mathbb{I}$, and defined by the time-ordered exponential,
\begin{equation}
t_2>t_1\,: \qquad U(t_2,t_1) = {\rm T}\,e^{-\frac{i}{\hbar}\int_{t_1}^{t_2}dt\,\hat{H}(t)}.
\end{equation}
In particular, for an arbitrary normalised quantum state, $\langle\psi,t|\psi,t\rangle=\langle \psi,t_0|\psi,t_0\rangle = 1$,
the quantum expectation value for the arbitrary observable $\hat{A}(t)$ is given by,
\begin{equation}
\langle\hat{A}\rangle(t)=\frac{\langle\psi,t |\hat{A}(t) |\psi,t\rangle}{\langle\psi,t |\psi,t\rangle}=
\langle \psi,t_0|U^\dagger(t,t_0)\,\hat{A}(t)\,U(t,t_0) |\psi,t_0\rangle.
\end{equation}
Within the Heisenberg picture time evolution of quantum observables is defined to be given by, say for the arbitrary observable $\hat{A}(t)$,
\begin{equation}
\hat{A}_H(t)=U^\dagger\,(t,t_0)\,\hat{A}(t)\,U(t,t_0),
\end{equation}
with thus in particular,
\begin{equation}
\hat{H}_H(t)=U^\dagger(t,t_0)\,\hat{H}(t)\,U(t,t_0).
\end{equation}
These operators are the solutions to the Heisenberg equation of motion in the Heisenberg picture, namely,
\begin{equation}
i\hbar\frac{d}{dt}\hat{A}_H(t)=i\hbar\frac{\partial}{\partial t}\hat{A}_H(t)\,+\,\left[\hat{A}_H(t),\hat{H}_H(t)\right],
\end{equation}
where explicitly,
\begin{equation}
\frac{\partial}{\partial t}\hat{A}_H(t) \equiv U^\dagger(t,t_0)\,\frac{\partial\hat{A}(t)}{\partial t}\,U(t,t_0).
\end{equation}
In particular in the case of a quantum dynamical Noether charge one has,
\begin{equation}
i\hbar\frac{d}{dt}\hat{Q}_H(t)=i\hbar\frac{\partial}{\partial t}\hat{Q}_H(t)\,+\,\left[\hat{Q}_H(t),\hat{H}_H(t)\right]=0,\qquad
\hat{Q}_H(t)=\hat{Q}_H(t_0)=\hat{Q}(t_0)=\hat{Q}_0.
\end{equation}
Consequently the quantum expectation value of a dynamical Noether charge for an arbitrary (normalised) quantum state is simply obtained in the form,
\begin{equation}
\langle\hat{Q}\rangle(t)=\langle\psi,t| \hat{Q}(t) |\psi,t\rangle=\langle\psi,t_0|\hat{Q}_H(t) |\psi,t_0\rangle
=\langle\psi,t_0| \hat{Q}_0 |\psi,t_0\rangle.
\end{equation}
In other words, under such general conditions, such a quantum expectation value indeed readily defines a time independent conserved
quantity whose value is set by the choice of initial conditions, on account of Noether's first theorem.

Quite many other possible choices of dynamical systems than the few simple illustrations presented herein,
for which similar considerations could be applied and developed to lead to further insight and understanding
into the classical and quantum dynamics easily come to one's mind. The celebrated Landau problem is certainly a case in point. However quantum field
theories, coupled to specific background spacetime geometries and with their infinite number of degrees of freedom and potentially huge dynamical
symmetries\cite{BG1}, offer great potential in that respect, as witnessed by the recent resurgence of interest into so-called
Bondi-Metzner-Sachs (BMS) symmetries (see for instance Ref.\cite{GC}).

\section*{Acknowledgements}

Data sharing is not applicable to this article as no new data were created or analysed in this study.

DBI acknowledges the support of an ``extraordinary" postdoctoral Fellowship of the
{\it Acad\'emie de Recherche et d'Enseignement Sup\'erieur} (ARES) of the Wallonia-Brussels
Federation of Belgium towards a six months stay at CP3 in 2019 during which the present work was
initiated. The work of JG is supported in part by the Institut Interuniversitaire des Sciences
Nucl\'eaires (IISN, Belgium).

\end{document}